\documentclass[review]{elsarticle}
\usepackage{lineno,hyperref}
\usepackage{amssymb,amsmath,latexsym}
\usepackage{mathrsfs}
\usepackage{tipa}
\usepackage{amsfonts}
\usepackage{graphicx}
\usepackage{subfigure}
\usepackage[title]{appendix}
\usepackage{color}
\usepackage{multirow}
\usepackage{amsbsy}
\usepackage{longtable}
\usepackage{floatrow}
\usepackage{subfigure}
\usepackage{indentfirst}
\usepackage{paralist}
\newtheorem{theorem}{Theorem}
\newtheorem{definition}{Definition}
\newtheorem{proposition}{Proposition}

\newtheorem{cor}{Corollary}
\usepackage[font=small]{caption}
\usepackage{algorithm}
\usepackage{algorithmic}
\modulolinenumbers[5]
\usepackage{xcolor}
\usepackage{hyperref} 
\usepackage{cases}
\usepackage{arydshln}

\hypersetup{               
	colorlinks=true,                
}

\AtBeginDocument{\hypersetup{pdfborder={0 0 1}}}

\usepackage[margin=2.5cm]{geometry}
\journal{}

\begin{document}
	
	\begin{frontmatter}
		
		\title{Block sparse signal recovery via minimizing the block $q$-ratio sparsity}
		
   	\fntext[myfootnote]{This work is supported by the Zhejiang Provincial Natural Science Foundation of China under Grant No. LQ21A010003.}
		
		\author[mymainaddress]{Zhiyong Zhou\corref{mycorrespondingauthor}}
		\cortext[mycorrespondingauthor]{Corresponding author}
		\ead{zhiyongzhou@zucc.edu.cn}
		
		
		\address[mymainaddress]{Department of Statistics, Zhejiang University City College, Hangzhou,
			310015, China}
		\begin{abstract}
		In this paper, we propose a method for block sparse signal recovery that minimizes the block $q$-ratio sparsity $\left(\lVert z\rVert_{2,1}/\lVert z\rVert_{2,q}\right)^{\frac{q}{q-1}}$ with $q\in[0,\infty]$. For the case of $1<q\leq\infty$, we present the theoretical analyses and the computing algorithms for both cases of the $\ell_2$-bounded and $\ell_{2,\infty}$-bounded noises. The corresponding unconstrained model is also investigated. Its superior performance in block sparse signal reconstruction is demonstrated by numerical experiments.
		\end{abstract}
		
		
	\end{frontmatter}

\section{Introduction}

The last two decades have seen increasing rapid advances in the field of compressive sensing (CS) (e.g., the monographs \cite{eldar2012compressed,foucart2013mathematical} and references therein). In the standard CS model $y=Ax+\varepsilon$, where $y\in\mathbb{R}^{m\times 1}$ is the vector of measurements, $A\in\mathbb{R}^{m\times N}$ is the pre-given measurement matrix, $x\in\mathbb{R}^N$ is the unknown signal, $\varepsilon$ is the measurement error, and the number of measurements is much less than the length of the signal (i.e., $m\ll N$), we aim to recover the unknown signal $x$ by using the under-determined measurements $y$ and the known matrix $A$. Research in this area has shown that, under the sparsity assumption of the signal, that is $x$ has only a few nonzero entries, and the measurement matrix $A$ is properly chosen (usually has some randomness), we can reliably recover $x$ from $y$ by certain algorithms, such as the following constrained $\ell_1$-minimization \cite{donoho2006compressed}:
\begin{align}
\min\limits_{z\in\mathbb{R}^N}\,\lVert z\rVert_1\quad \text{subject to \quad $\lVert y-Az\rVert_2\leq \eta $}.
\end{align}
Meanwhile, to gain better recovery performances, various non-convex algorithms have been proposed, including $\ell_p$ ($0<p<1$) \cite{chartrand2008restricted,foucart2009sparsest}, $\ell_1-\ell_2$ \cite{yin2015minimization}, transformed $\ell_1$ (TL1) \cite{zhang2018minimization}, smoothly clipped absolute deviation (SCAD) \cite{fan2001variable}, minimax concave penalty (MCP) \cite{zhang2010nearly}, and $\ell_1/\ell_2$ \cite{rahimi2018scale,wang2020accelerated}, among others. As a direct extension of the $\ell_1/\ell_2$ method, very recently \cite{zhou2020minimization} proposed a more general scale invariant approach for sparse recovery via minimizing the $q$-ratio sparsity measure $s_q(z)=(\lVert z\rVert_1/\lVert z\rVert_q)^{q/(q-1)}$ with $q\in[0,\infty]$. However, previous published studies on this kind of scale invariant approaches such as \cite{rahimi2018scale,wang2020accelerated,zhou2020minimization} are limited to the case of non-block sparse signal recovery. The present paper sets out to investigate the minimization of the block $q$-ratio sparsity measure given in \cite{zhou2017estimation} for block sparse signal recovery.

When the nonzero entries of a sparse signal occur in clusters, we use block sparsity to characterize this additional structure. There are a lot of studies on the block sparse model, both on its wide range of practical applications \cite{majumdar2010compressed,mishali2009blind,parvaresh2008recovering} and on its theoretical analysis results \cite{blumensath2009sampling,chen2006theoretical,eldar2010block,eldar2009robust,elhamifar2012block}. Suppose $N=\sum_{j=1}^{M}d_j$, then the $j$-th block of a length-$N$ vector $x$ over $\mathcal{I}=\{d_1,\cdots,d_M\}$ is denoted by $x[j]$. That means the $j$-th block is of length $d_j$, and the blocks are formed sequentially as follows: 
\begin{align}
x=(\underbrace{x_1\cdots x_{d_1}}_{x^{T}[1]}\underbrace{x_{d_1+1}\cdots x_{d_1+d_2}}_{x^{T}[2]}\cdots\underbrace{x_{N-d_M+1}\cdots x_N}_{x^{T}[M]})^T. \label{signal}
\end{align}
Without loss of generality, for simplicity we may take $d_1=d_2=\cdots=d_M=d$ so that $N=Md$. Based on this setting, a vector $x\in\mathbb{R}^N$ is called block $k$-sparse if it has at most $k$ non-zero blocks. In other words, we have $\lVert x\rVert_{2,0}=\sum_{j=1}^{M}I(\lVert x[j]\rVert_2\neq 0)\leq k$ for any block $k$-sparse vector $x$.  

The corresponding extended versions of sparse algorithms have been developed to reconstruct block sparse signal, such as the mixed $\ell_2/\ell_1$ norm recovery algorithm given in \cite{eldar2010block}:\begin{align}
\min\limits_{z\in\mathbb{R}^N}\,\lVert z\rVert_{2,1}\quad \text{subject to \quad $\lVert y-Az\rVert_2\leq \eta $}, \label{mixedl2l1}
\end{align}
where $\lVert z\rVert_{2,1}=\sum_{j=1}^M \lVert z[j]\rVert_2$. The mixed $\ell_2/\ell_1$ method is the block version of the $\ell_1$-minimization method, while the block version of the non-convex $\ell_p$ ($0<p<1$) method is the mixed $\ell_2/\ell_p$ method \cite{wang2013recovery,wang2014restricted} by solving \begin{align}
\min\limits_{z\in\mathbb{R}^N}\,\lVert z\rVert_{2,p}^p\quad \text{subject to \quad $\lVert y-Az\rVert_2\leq \eta $},
\end{align}
with $\lVert z\rVert_{2,p}=(\sum_{j=1}^M \lVert z[j]\rVert_2^p)^{1/p}$. Other typical algorithms for block sparse recovery include the mixed $\ell_q/\ell_1$ ($q\geq 1$) norm recovery algorithm \cite{elhamifar2012block}, group lasso \cite{yuan2006model}, iterative reweighted $\ell_2/\ell_1$ recovery algorithms \cite{zeinalkhani2015iterative}, the $\ell_2/\ell_{1-2}$ method (the block version of $\ell_1-\ell_2$ via the minimization of $\lVert \cdot\rVert_{2,1}-\lVert \cdot\rVert_2$) \cite{wang2017block}, the block version of Orthogonal Matching Pursuit (OMP) algorithm \cite{eldar2010block} and the extensions of the Compressive Sampling Matching Pursuit (CoSaMP) algorithm and of the Iterative Hard Thresholding (IHT) to the model-based CS \cite{baraniuk2010model}, which includes block sparse model as a special case. 

This work is inspired by \cite{zhou2020minimization}, in which a $q$-ratio sparsity minimization based method was proposed for non-block sparse signal recovery. The benefit of this novel method is that it enjoys a superior performance when highly coherent measurement matrices are confronted. In the present paper, we extend this method to the framework of block sparse signal recovery via minimizing the block version of $q$-ratio sparsity given in \cite{zhou2017estimation} (namely the block $q$-ratio sparsity). Our main contributions are three folds: \begin{enumerate}
	\item We propose the minimization of the block $q$-ratio sparsity for block sparse signal recovery, which extends our previous work \cite{zhou2020minimization} from non-block case to block case.
	\item We consider both the $\ell_2$-bounded and $\ell_{2,\infty}$-bounded noise cases, and obtain the stable and robust recovery results in terms of $q$-ratio block constrained minimal singular value (BCMSV). What's more, the theoretical analyses for the unconstrained-version model are also established.
	\item We present the block version of convex-concave procedure algorithm given in \cite{zhou2020minimization}  and conduct numerical experiments to show its good performances.
\end{enumerate}

\subsection{Organization and Notations}

The overall structure of this paper takes the form of six sections. In Section 2, we present the definition of block $q$-ratio sparsity and propose the block sparse signal recovery methodology via minimizing the block $q$-ratio sparsity. In Section 3, we provide a verifiable sufficient condition for the exact block sparse recovery and derive the reconstruction error bounds based on $q$-ratio BCMSV for the proposed method in the case of $1<q\leq \infty$, involving both constrained and unconstrained models. In Section 4, we design algorithms to solve the problem. Section 5 contains the numerical experiments. Finally, conclusions are included in Section 6. 

Throughout the paper, we introduce the notations $[M]$ for the block index set $\{1,2,\cdots,M\}$ and $|S|$ for the cardinality of a block index subset $S\subseteq [M]$. We write $S^c$ for the complement $[M]\setminus S$ of a set $S$ in $[M]$. The block support of a vector $x\in\mathbb{R}^N$ is the index set of its nonzero blocks, i.e., $\mathrm{bsupp}(x):= \{j \in [M]: \lVert x[j]\rVert_2\neq 0\}$. The mixed $\ell_2/\ell_q$-norm $\lVert x\rVert_{2,q}=(\sum_{j=1}^M \lVert x[j]\rVert_2^q)^{1/q}$ for any $q\in(0,\infty)$, while $\lVert x\rVert_{2,\infty}=\max_{1\leq j\leq M} \lVert x[j]\rVert_2$. For a vector $x\in\mathbb{R}^N$ and a block index subset $S\subseteq [M]$, $x_S$ will denote the vector equal to $x$ on the block index set $S$ and zero elsewhere.  

\section{Minimization of the block $q$-ratio sparsity}

The traditional block sparsity measure $\lVert \cdot\rVert_{2,0}$ has a severe practical drawback of
being not sensitive to blocks with small $\ell_2$ norm. As a soft version, the entropy based block sparsity measure named block $q$-ratio sparsity was proposed in \cite{zhou2017estimation}, which possesses many nice properties including continuity, scale-invariance, non-increasing with respect to $q$ and range equal to $[1,M]$. For more detailed arguments about this block sparsity measure, readers can refer to \cite{zhou2017estimation}. To be self-contained, here we give the full definition of block $q$-ratio sparsity. 

\begin{definition}(\cite{zhou2017estimation})
	For any non-zero $z\in\mathbb{R}^N$ and non-negative $q\notin\{0,1,\infty\}$, the block $q$-ratio sparsity level of $z$ is defined as \begin{align}
	k_{q}(z)=\left(\frac{\lVert z\rVert_{2,1}}{\lVert z\rVert_{2,q}}\right)^{\frac{q}{q-1}}.
	\end{align} 
    The cases of $q\in\{0,1,\infty\}$ are evaluated as limits: 
	$k_0(z)=\lim\limits_{q\rightarrow 0} k_q(z)=\lVert z\rVert_{2,0}$, $k_1(z)=\lim\limits_{q\rightarrow 1} k_q(z)=\exp\left(-\sum_{j=1}^M \frac{\lVert z[j]\rVert_2}{\lVert z\rVert_{2,1}}\ln \frac{\lVert z[j]\rVert_2}{\lVert z\rVert_{2,1}}\right)$, $k_\infty(z)=\lim\limits_{q\rightarrow \infty} k_q(z)=\frac{\lVert z\rVert_{2,1}}{\lVert z \rVert_{2,\infty}}$.
\end{definition}

Based on this soft block sparsity measure, in this paper we propose the following non-convex minimization problems for block sparse signal recovery: 
\begin{subnumcases}{\min\limits_{z\in\mathbb{R}^N} k_q(z)\quad \text{subject to}\quad}
\lVert y-Az\rVert_2\leq \eta, \\
\lVert A^{T}(y-Az)\rVert_{2,\infty}\leq \mu,
\end{subnumcases}
where $y=Ax+\varepsilon$ with $\lVert \varepsilon\rVert_2\leq \eta$ or $\lVert \varepsilon\rVert_{2,\infty}\leq \mu$, and some $q\in[0,\infty]$ is pre-given. Here we consider both cases of the $\ell_2$-bounded and $\ell_{2,\infty}$-bounded noises.

\begin{figure}[htbp]
	\centering
	\includegraphics[width=\textwidth,height=0.4\textheight]{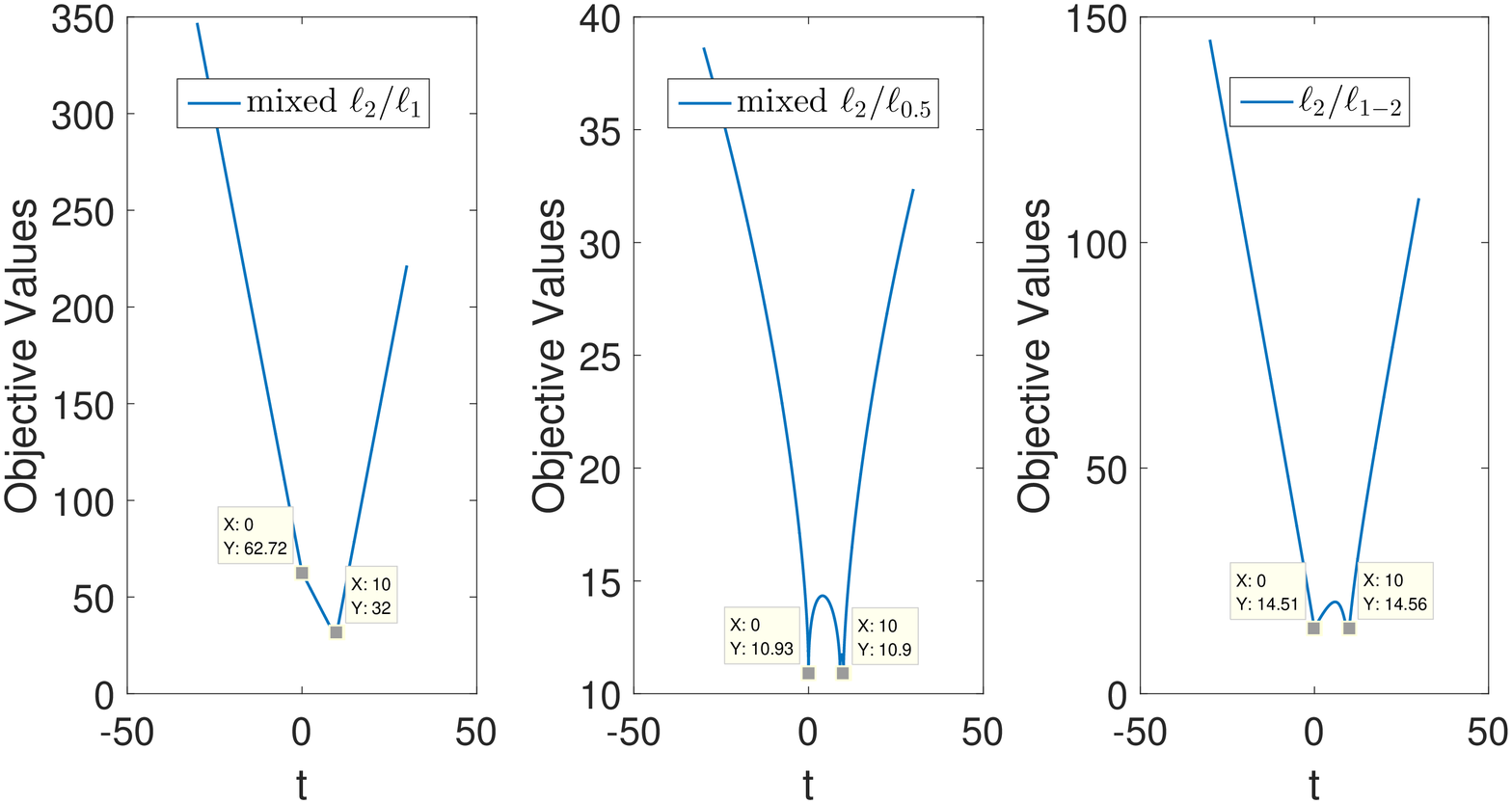} 
	\caption{The objective functions of a toy example for the mixed $\ell_2/\ell_1$, the mixed $\ell_2/\ell_{0.5}$ and the $\ell_2/\ell_{1-2}$.} \label{toy_examplemixed}
\end{figure}

\begin{figure}[htbp]
	\centering
	\includegraphics[width=\textwidth,height=0.4\textheight]{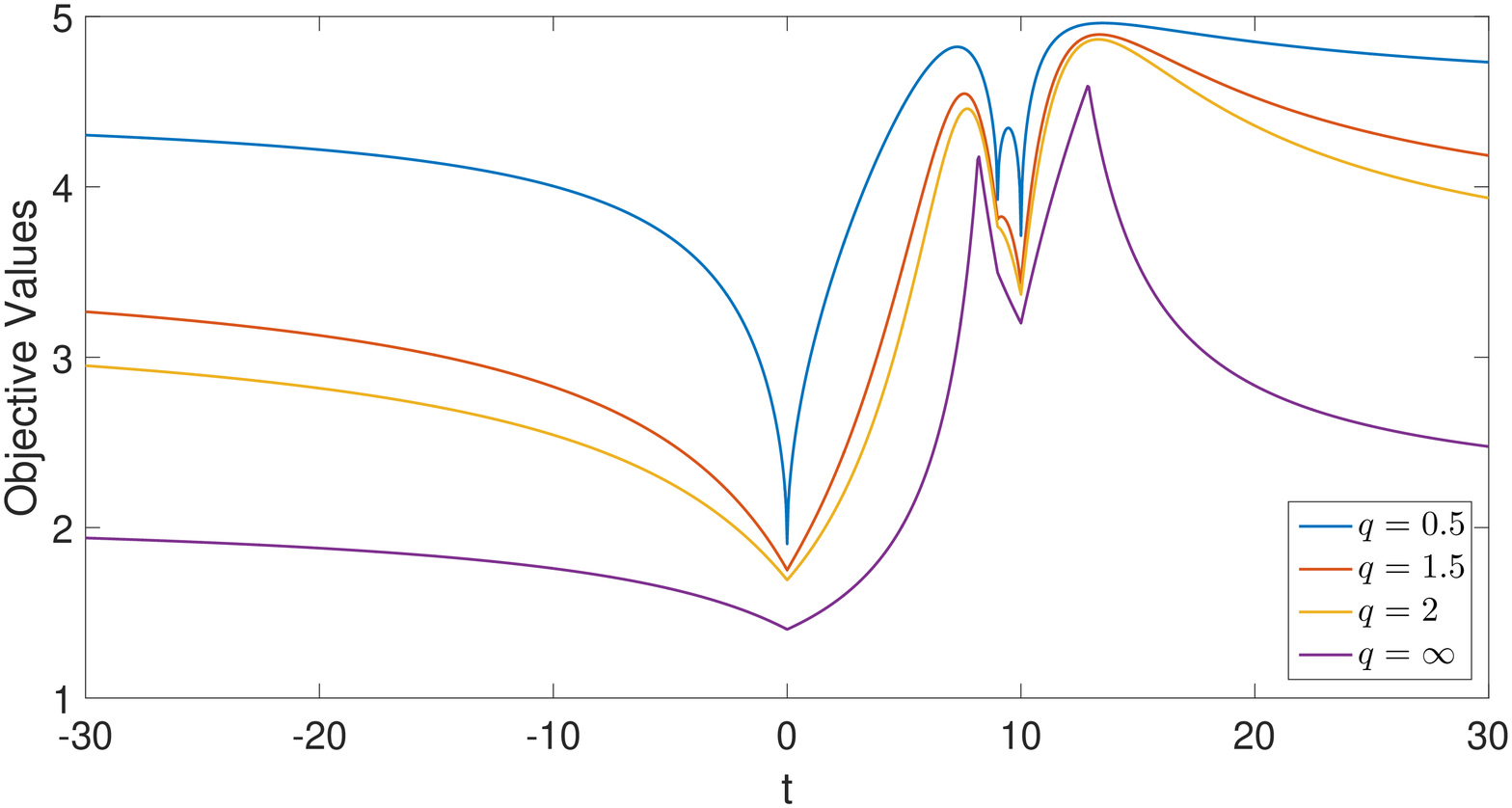} 
	\caption{The objective functions of a toy example used to illustrate that minimizing the block $q$-ratio sparsity $k_q(\cdot)$ can find $t=0$ as the global minimizer.} \label{toy_example}
\end{figure}

In order to illustrate the block sparsity promoting ability of the block $q$-ratio sparsity minimization problem, we revisit a toy example previously discussed in \cite{rahimi2018scale,zhou2020minimization}. Specifically, we let the measurement matrix \[A=\left(\begin{array}{c:c:c:cc:c}
1& -1&  0& 0&  0&0 \\
1&  0& -1& 0&  0&0 \\
0& 1& 1& 1& 0& 0 \\
2& 2& 0& 0& 1& 0 \\
1& 1& 0& 0& 0& -1
\end{array}
\right)
\in\mathbb{R}^{5\times 6},\] and the measurement vector $y=(0,0,20,40,18)^T\in\mathbb{R}^5$. Then, it is straightforward to show that any solution of $Az=y$ has the form of $z=(t,t,t,20-2t,40-4t,2(t-9))^T$ for some $t\in\mathbb{R}$.  In this case we assume the block sizes go as that $d_1=d_2=d_3=1, d_4=2, d_5=1$. It is easy to notice that the block sparsest solution occurs at $t=0$, where its block sparsity is 2. Other local solutions include $t=10$ and $t=9$ with block sparsity being 4. As can be seen in Figure \ref{toy_examplemixed}, among the methods mentioned (including mixed $\ell_2/\ell_1$, mixed $\ell_2/\ell_{0.5}$ and $\ell_2/\ell_{1-2}$), only $\ell_2/\ell_{1-2}$ model can find the global minimizer $t=0$. Moreover, according to the result displayed in Figure \ref{toy_example}, our proposed methods with varying choices of $q$ are all able to find the global minimizer at $t=0$. Looking at Figure \ref{toy_example}, it is apparent that the objective functions have two local minimizers ($t=0$ and $t=10$) when $q=1.5,2,\infty$, while it has three local minimizers ($t=0$, $t=9$ and $t=10$) when $q=0.5$. This provides evidence that it is much harder to solve the minimization of block $q$-ratio sparsity for the case of $0<q\leq 1$ than for the case of $1<q\leq \infty$.

\begin{figure}[htbp]
	\centering
	\begin{minipage}[t]{0.49\textwidth}
		\includegraphics[width=\textwidth,height=0.36\textheight,keepaspectratio]{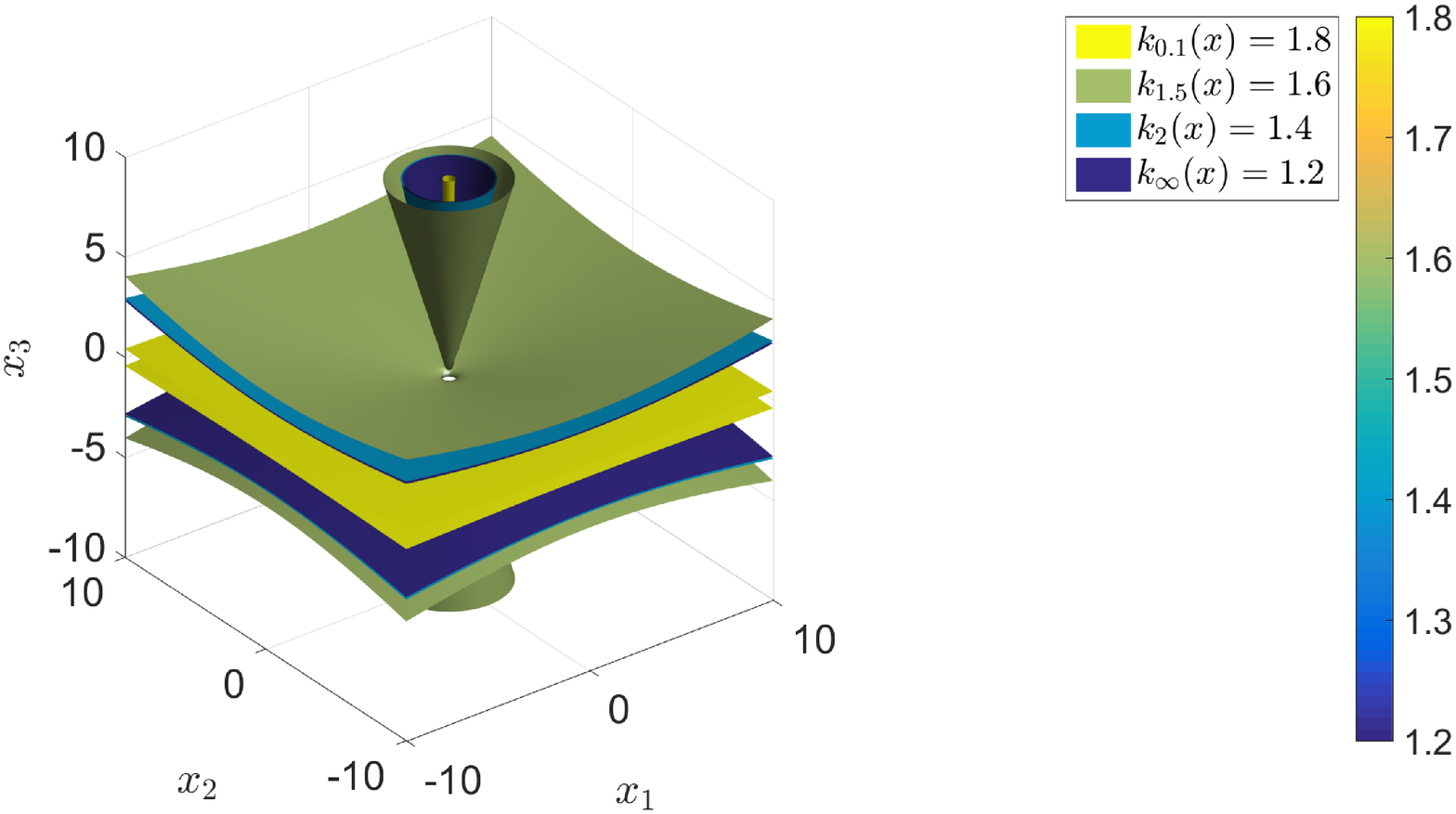}
	\end{minipage}
	\begin{minipage}[t]{0.49\textwidth}
		\includegraphics[width=\textwidth,height=0.36\textheight,keepaspectratio]{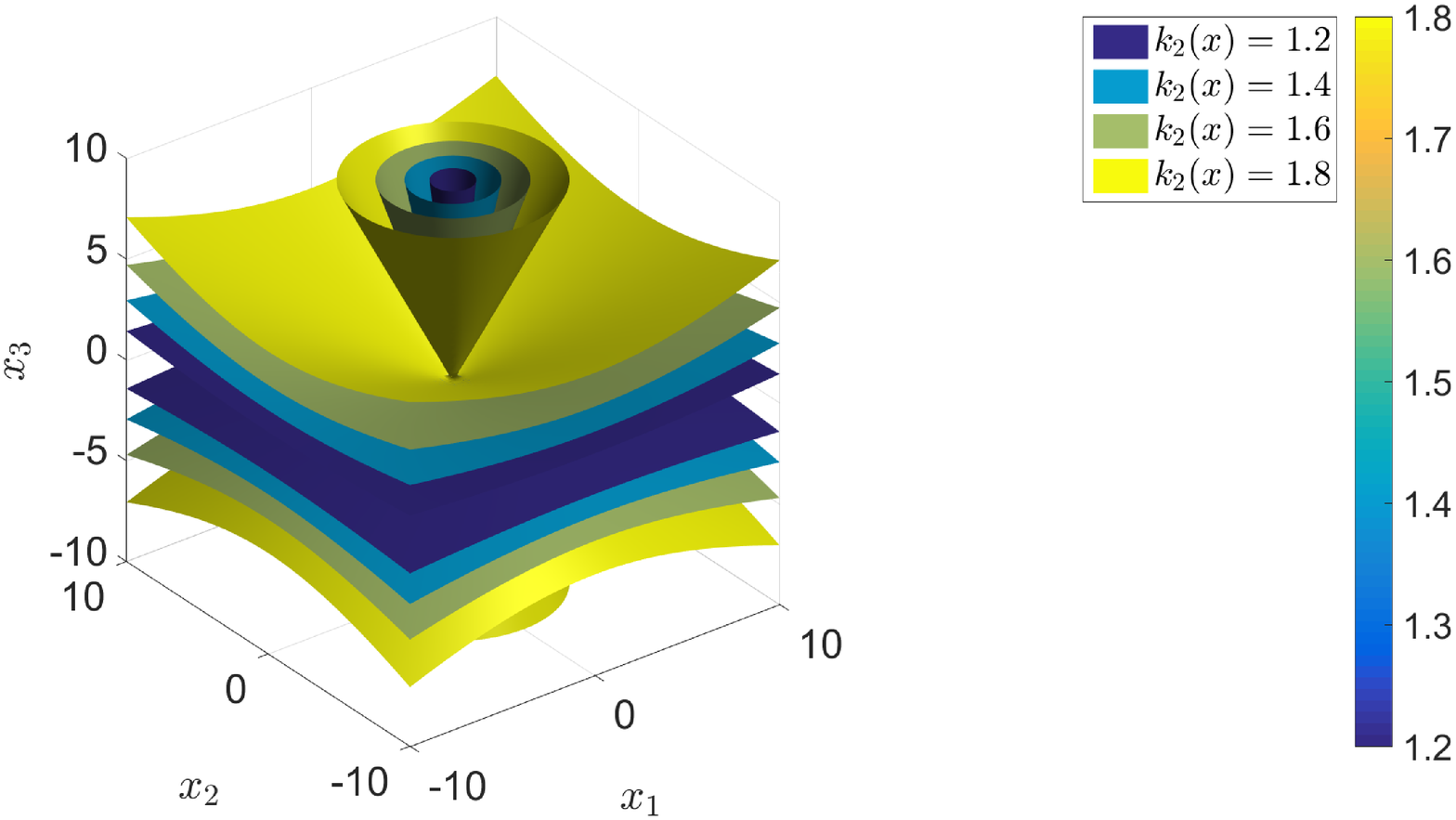}
	\end{minipage}
	\caption{The isosurface plots of the block $q$-ratio sparsity $k_q(x)$ for $x=(x_1,x_2,x_3)^T\in\mathbb{R}^3$, where $x$ has two blocks with $x[1]=(x_1,x_2)^T$ and $x[2]=x_3$.} \label{isosurface}
\end{figure}

On the other hand, we present the isosurface plots for the block $q$-ratio sparsity $k_q(x)$ of $x\in\mathbb{R}^3$ with different values of $q$. As shown in Figure \ref{isosurface}, similar non-convex patterns arise while varying $q$ from $0.1,1.5,2$ and $\infty$. The fact that the isosurface of $k_2(x)$ approaches the planes $x_3=0$ and $x_1=x_2=0$ as its value gets small reflects its ability to promote block sparsity. Meanwhile, the sparsity-promoting analysis technique used in \cite{huang2018sparse} can also be adopted here to show that minimizing the block $q$-ratio sparsity in an orthant of the Euclidean space $\mathbb{R}^N$ leads to solutions on the boundary, i.e., block sparser solutions. And it can be shown that minimizing the block $q$-ratio sparsity has the energy-promoting property, namely it promotes high-energy blocks while suppressing the rest low-energy blocks, see \cite{huang2018sparse} for detailed discussions.

As done in \cite{zhou2020minimization}, in this present paper we merely focus on the minimization problems with pre-given $q\in (1,\infty]$, in which case they are equivalent to solve the constrained $\ell_{2,1}/\ell_{2,q}$ minimization problems:
\begin{subnumcases}{\min\limits_{z\in\mathbb{R}^N} \frac{\lVert z\rVert_{2,1}}{\lVert z\rVert_{2,q}}\quad \text{subject to} \quad}
\lVert y-Az\rVert_2\leq \eta \label{blocknorm_ratio}, \\
\lVert A^{T}(y-Az)\rVert_{2,\infty}\leq \mu. \label{DS}
\end{subnumcases}

\section{Recovery analysis}

The section below studies the global optimality results for the $\ell_{2,1}/\ell_{2,q}$ minimization with $q\in(1,\infty]$. We firstly establish a sufficient condition for the exact block sparse recovery using the $\ell_{2,1}/\ell_{2,q}$ minimization with $q\in(1,\infty]$. For some pre-given $q\in(1,\infty]$, we discuss the noiseless $\ell_{2,1}/\ell_{2,q}$ minimization problem:\begin{align}
\min\limits_{z\in\mathbb{R}^N} \frac{\lVert z\rVert_{2,1}}{\lVert z\rVert_{2,q}}\quad \text{subject to \quad $Az=Ax$}.\label{noiseless}
\end{align}
It can be easily verified that the sufficient and necessary condition for exactly recovering the block $k$-sparse $x$ via (\ref{noiseless}) is given by the following null space property \cite{cohen2009compressed}: \begin{align}
\frac{\lVert x\rVert_{2,1}}{\lVert x\rVert_{2,q}}<\frac{\lVert x+h\rVert_{2,1}}{\lVert x+h\rVert_{2,q}},\quad \forall\,\, h\in \mathrm{ker}(A)\setminus \{0\}. \label{nsp}
\end{align}
As a consequence, we are able to obtain the following verifiable sufficient condition that guarantees the uniform exact block sparse recovery using the noiseless $\ell_{2,1}/\ell_{2,q}$ problem (\ref{noiseless}). It acts as a direct extension of Proposition 3 in \cite{zhou2020minimization}.

\begin{proposition}
	For some pre-given $q\in(1,\infty]$, if $x$ is block $k$-sparse such that \begin{align}
	k<\inf\limits_{h\in \mathrm{ker}(A)\setminus \{0\}} 3^{\frac{q}{1-q}}k_q(h), \label{sufficient}
	\end{align}
	then the unique solution to the problem (\ref{noiseless}) is the truth $x$.
\end{proposition}

\noindent
{\bf Proof.}  The proof of this proposition is almost identical to the proof of Proposition 3 in \cite{zhou2020minimization}, with the major change being the substitution of the non-block norms for block norms. The proof is reproduced here for the sake of completeness. To prove the result, it suffices to verify the null space property (\ref{nsp}) mentioned above. We assume that the block support of the block $k$-sparse $x$ is $\mathrm{bsupp}(x)=S$ such that $|S|\leq k$. For any  $q\in(1,\infty]$ and $h\in \mathrm{ker}(A)\setminus \{0\}$, it holds that
\begin{align*}\frac{\lVert x+h\rVert_{2,1}}{\lVert x+h\rVert_{2,q}} \geq \frac{\lVert x\rVert_{2,1}+\lVert h\rVert_{2,1}-2\lVert h_S\rVert_{2,1}}{\lVert x\rVert_{2,q}+\lVert h\rVert_{2,q}} \geq \min\left\{\frac{\lVert x\rVert_{2,1}}{\lVert x\rVert_{2,q}},\frac{\lVert h\rVert_{2,1}-2\lVert h_S\rVert_{2,1}}{\lVert h\rVert_{2,q}}\right\}.
\end{align*}
where we adopt the facts that $\lVert x+h\rVert_{2,1}=\lVert x+h_S+h_{S^c}\rVert_{2,1}\geq \lVert x\rVert_{2,1}+\lVert h_{S^c}\rVert_{2,1}-\lVert h_S\rVert_{2,1}=\lVert x\rVert_{2,1}+\lVert h\rVert_{2,1}-2\lVert h_S\rVert_{2,1}$ and $\lVert x+h\rVert_{2,q}\leq \lVert x\rVert_{2,q}+\lVert h\rVert_{2,q}$. 

If $k<3^{\frac{q}{1-q}}k_q(h)$, then we obtain that $\frac{\lVert h\rVert_{2,1}}{\lVert h\rVert_{2,q}}>3k^{1-1/q}>k^{1-1/q}+2\frac{\lVert h_S\rVert_{2,1}}{\lVert h\rVert_{2,q}}$, which leads to $\frac{\lVert h\rVert_{2,1}-2\lVert h_S\rVert_{2,1}}{\lVert h\rVert_{2,q}}>k^{1-1/q}\geq\frac{\lVert x\rVert_{2,1}}{\lVert x\rVert_{2,q}}$. Therefore, the null space property (\ref{nsp}) holds and the proof is completed. \\

What follows is the stable and robust recovery analysis results for the $\ell_{2,1}/\ell_{2,q}$ minimization problems involving both constrained and unconstrained models. We start with the definition of $q$-ratio block constrained minimal singular values (BCMSV), which is a computable quality measure for the measurement matrix. It is a block version of the $q$-ratio constrained minimal singular values (CMSV) proposed and systematically studied in \cite{zhou2018q,zhou2018sparse}. As an efficient theoretical analysis tool for block sparse recovery, $q$-ratio BCMSV has been successfully used in establishing reconstruction error bounds for the block basis pursuit (BBP), the block Dantzig selector (BDS), and the group lasso, see \cite{wang2019error} for detailed arguments.

\begin{definition}
	For any real number $s\in[1,M]$, $q\in (1, \infty]$ and matrix $A\in\mathbb{R}^{m\times N}$, the $q$-ratio block constrained minimal singular value (BCMSV) of $A$ is defined as \begin{align}
	\beta_{q,s}(A)=\min\limits_{z\neq 0,k_q(z)\leq s}\,\,\frac{\lVert Az\rVert_2}{\lVert z\rVert_{2,q}}. \label{bcmsv}
	\end{align}
\end{definition}

\subsection{Constrained Models}

Let us now turn to the recovery analysis results for the constrained models (\ref{blocknorm_ratio}) and (\ref{DS}) based on the $q$-ratio BCMSV. In addition to cover the main results in \cite{zhou2020minimization} for the non-block sparse recovery with $\ell_2$-bounded noise, this subsection also considers the $\ell_{2,\infty}$-bounded noise case. The corresponding results for the case that the true signal $x$ is exactly block sparse are list as follows. 

\begin{theorem}
	Suppose $x$ is non-zero and block $k$-sparse. For any $1<q\leq \infty$ and $\beta_{q,3^{\frac{q}{q-1}}k}(A)>0$, 
	
	\noindent
	(1) If the noise in (\ref{blocknorm_ratio}) satisfies $\lVert \varepsilon \rVert_2\leq \eta$, then the solution $\hat{x}$ to the problem (\ref{blocknorm_ratio}) obeys \begin{align}
	\lVert \hat{x}-x\rVert_{2,q} &\leq \frac{2\eta}{\beta_{q,3^{\frac{q}{q-1}}k}(A)},  \\
	\lVert \hat{x}-x\rVert_{2,1} &\leq  \frac{6 k^{1-1/q}\eta}{\beta_{q,3^{\frac{q}{q-1}}k}(A)}. 
	\end{align}
	\noindent
	(2) If the noise in (\ref{DS}) satisfies $\lVert A^{T}\varepsilon \rVert_{2,\infty}\leq \mu$, then the solution $\hat{x}$ to the problem (\ref{DS}) obeys \begin{align}
	\lVert \hat{x}-x\rVert_{2,q} &\leq \frac{6k^{1-1/q}\mu}{\beta_{q,3^{\frac{q}{q-1}}k}^2(A)}, \\
	\lVert \hat{x}-x\rVert_{2,1} &\leq  \frac{18 k^{2-2/q}\mu}{\beta_{q,3^{\frac{q}{q-1}}k}^2(A)}.
	\end{align}
\end{theorem}

\noindent \\
{\bf Proof.} As $x$ is block $k$-sparse, let us assume that $\mathrm{bsupp}(x)=S$ and $|S|\leq k$. For both of the constrained models (\ref{blocknorm_ratio}) and (\ref{DS}), we denote the residual by $h:=\hat{x}-x$. Due to $\hat{x}=x+h$ is the minimum among all $z$ satisfying the constraints of the models (\ref{blocknorm_ratio}) and (\ref{DS}), it follows that \begin{align*}
\frac{\lVert x+h\rVert_{2,1}}{\lVert x+h\rVert_{2,q}}\leq \frac{\lVert x\rVert_{2,1}}{\lVert x\rVert_{2,q}},
\end{align*}
which leads to \begin{align}
\lVert x+h\rVert_{2,1}\cdot\lVert x\rVert_{2,q}\leq \lVert x\rVert_{2,1}\cdot\lVert x+h\rVert_{2,q}. \label{ineq}
\end{align}
Meanwhile, we have \begin{align*}
\lVert x+h\rVert_{2,1}=\lVert x_S+h_S\rVert_{2,1}+\lVert x_{S^c}+h_{S^c}\rVert_{2,1}\geq \lVert x_S\rVert_{2,1}-\lVert h_S\rVert_{2,1}+\lVert h_{S^c}\rVert_{2,1}=\lVert x\rVert_{2,1}-\lVert h_S\rVert_{2,1}+\lVert h_{S^c}\rVert_{2,1},
\end{align*}
and $\lVert x+h\rVert_{2,q}\leq \lVert x\rVert_{2,q}+\lVert h\rVert_{2,q}$. Consequently, we infer that \begin{align*}
(\lVert x\rVert_{2,1}-\lVert h_S\rVert_{2,1}+\lVert h_{S^c}\rVert_{2,1})\cdot\lVert x\rVert_{2,q}\leq \lVert x\rVert_{2,1}\cdot (\lVert x\rVert_{2,q}+\lVert h\rVert_{2,q}).
\end{align*}
Then it holds that \begin{align}
\lVert h_{S^c}\rVert_{2,1}\leq \lVert h_S\rVert_{2,1}+\frac{\lVert x\rVert_{2,1}}{\lVert x\rVert_{2,q}}\lVert h\rVert_{2,q}=\lVert h_S\rVert_{2,1}+k_q(x)^{1-1/q}\lVert h\rVert_{2,q},
\end{align}
which implies that \begin{align}
\lVert h\rVert_{2,1}=\lVert h_S\rVert_{2,1}+\lVert h_{S^c}\rVert_{2,1}\leq 2\lVert h_S\rVert_{2,1}+k_q(x)^{1-1/q}\lVert h\rVert_{2,q}\leq (2k^{1-1/q}+k_q(x)^{1-1/q})\lVert h\rVert_{2,q}.
\end{align}
Thus we arrive at the conclusion that for any $1<q\leq \infty$, $k_q(h)=\left(\frac{\lVert h\rVert_{2,1}}{\lVert h\rVert_{2,q}}\right)^{\frac{q}{q-1}}\leq \left(2k^{1-1/q}+k_q(x)^{1-1/q}\right)^{\frac{q}{q-1}}\leq 3^{\frac{q}{q-1}} k$, by adopting $k_q(x)\leq k_0(x)=\lVert x\rVert_{2,0}\leq k$. 

(1) As for the problem (\ref{blocknorm_ratio}), because $\hat{x}$ satisfies the constraint $\lVert y-A\hat{x}\rVert_2\leq \eta$ and $\lVert y-Ax\rVert_2=\lVert \varepsilon\rVert_2\leq \eta$, it follows that \begin{align}
\lVert Ah\rVert_2=\lVert A(\hat{x}-x)\rVert_2\leq \lVert A\hat{x}-y\Vert_2+\lVert y-Ax\rVert_2\leq 2\eta. \label{Ah_bound}
\end{align}
Then, according to the definition of $q$-ratio BCMSV and $k_q(h)\leq 3^{\frac{q}{q-1}}k$, it holds that \begin{align*}
\beta_{q,3^{\frac{q}{q-1}}k}(A)\lVert h\rVert_{2,q}\leq \lVert Ah\rVert_2\leq 2\eta\Rightarrow \lVert h\rVert_{2,q}\leq \frac{2\eta}{\beta_{q,3^{\frac{q}{q-1}}k}(A)}.
\end{align*}
Meanwhile, $\lVert h\rVert_{2,1}\leq 3k^{1-1/q}\lVert h\rVert_{2,q}\Rightarrow \lVert h\rVert_{2,1}\leq \frac{6k^{1-1/q}\eta}{\beta_{q,3^{\frac{q}{q-1}}k}(A)}$. This completes the proof of results for the problem (\ref{blocknorm_ratio}).

(2) With regard to the problem (\ref{DS}), since $\lVert A^{T}\varepsilon\rVert_{2,\infty}\leq \mu$, we have \begin{align}
\lVert A^{T}Ah\rVert_{2,\infty}\leq \lVert A^{T}(y-A\hat{x})\rVert_{2,\infty}+\lVert A^{T}(y-Ax)\rVert_{2,\infty}\leq 2\mu.
\end{align}
Therefore, \begin{align*}
\lVert Ah\rVert_2^2=\langle Ah, Ah\rangle=\langle h, A^{T}Ah\rangle\leq \lVert h\rVert_{2,1}\lVert A^{T}Ah\rVert_{2,\infty}\leq 2\mu\lVert h\rVert_{2,1}.
\end{align*}
Thus, together with $k_q(h)\leq 3^{\frac{q}{q-1}}k$, we obtain that \begin{align*}
\beta_{q,3^{\frac{q}{q-1}}k}^2(A)\lVert h\rVert_{2,q}^2\leq \lVert Ah\rVert_2^2\leq 2\mu\lVert h\rVert_{2,1}\leq 6k^{1-1/q}\mu\lVert h\rVert_{2,q},
\end{align*}
which leads to \begin{align}
\lVert h\rVert_{2,q}\leq \frac{6k^{1-1/q}\mu}{\beta_{q,3^{\frac{q}{q-1}}k}^2(A)}.
\end{align}
Hence, $\lVert h\rVert_{2,1}\leq 3 k^{1-1/q}\lVert h\rVert_{2,q}\leq \frac{18k^{2-2/q}\mu}{\beta_{q,3^{\frac{q}{q-1}}k}^2(A)}$. We have thus proved the theorem.

\noindent\\
{\bf Remark.} As studied in the Theorem 3 of \cite{wang2019error}, this sort of $q$-ratio BCMSV based condition $\beta_{q,3^{\frac{q}{q-1}}k}(A)>0$ is fulfilled with high probability for subgaussian random matrix when the number of its measurements is reasonably large compared to the block sparsity level $k$. And for any pre-given measurement matrix $A$, its $q$-ratio BCMSV can be computed approximately so that the concise error bounds established in this theorem can be well computed. \\

The following corollary follows immediately from Theorem 1 by letting $\eta=0$ in  (\ref{blocknorm_ratio}) or $\mu=0$ in  (\ref{DS}). The sufficient condition that $\beta_{q,3^{\frac{q}{q-1}}k}(A)>0$ for a perfect block sparse recovery via the noiseless $\ell_{2,1}/\ell_{2,q}$ minimization presented here  is a bit stronger than the condition that $\beta_{q,2^{\frac{q}{q-1}}k}(A)>0$ given for the mixed $\ell_{2}/\ell_1$-minimization in \cite{wang2019error}.

\begin{cor}
	For any non-zero block $k$-sparse signal $x$ and any $q\in(1,\infty]$, if the condition $\beta_{q,3^{\frac{q}{q-1}}k}(A)>0$ holds, then the unique solution of (\ref{noiseless}) is exactly the truth $x$. 
\end{cor}

Having analyzed the case that the true signal is exactly block sparse in detail, we now move on to consider the case that it is block compressible, i.e., it can be well approximated by an exactly block sparse signal. For any $x\in\mathbb{R}^N$, throughout this paper we denote by $x^{k}$ its best block $k$-sparse approximation with respective to $\lVert\cdot\rVert_{2,1}$ (i.e., $x^{k}=\mathop{\arg\min}\limits_{\lVert u\rVert_{2,0}\leq k}\lVert x-u\rVert_{2,1}$).

\begin{theorem}

Let non-zero $x\in\mathbb{R}^N$ and denote $B_q(k,x)=(4k^{1-1/q}+k_q(x)^{1-1/q})^{\frac{q}{q-1}}$. For any $1<q\leq \infty$ and $\beta_{q,B_q(k,x)}(A)>0$,
	
	\noindent
	(1) If the noise in (\ref{blocknorm_ratio}) satisfies $\lVert \varepsilon \rVert_2\leq \eta$, then the solution $\hat{x}$ to the problem (\ref{blocknorm_ratio}) obeys
	 \begin{align}
	\lVert\hat{x}-x\rVert_{2,q}&\leq \frac{2\eta}{\beta_{q,B_q(k,x)}(A)}+k^{1/q-1}\lVert x-x^{k}\rVert_{2,1}, \label{robust} \\
	\lVert\hat{x}-x\rVert_{2,1}&\leq \frac{(4k^{1-1/q}+2k_q(x)^{1-1/q})\eta}{\beta_{q,B_q(k,x)}(A)}+(4+(k_q(x)/k)^{1-1/q})\lVert x-x^{k}\rVert_{2,1}. \label{robustl1}
	\end{align}
	\noindent
	(2) If the noise in (\ref{DS}) satisfies $\lVert A^{T}\varepsilon \rVert_{2,\infty}\leq \mu$, then the solution $\hat{x}$ to the problem (\ref{DS}) obeys
	 \begin{align}
	\lVert\hat{x}-x\rVert_{2,q}&\leq \frac{2B_q(k,x)^{1-1/q}\mu}{\beta_{q,B_q(k,x)}^2(A)}+k^{1/q-1}\lVert x-x^{k}\rVert_{2,1}, \label{DSrobust} \\
	\lVert\hat{x}-x\rVert_{2,1}&\leq \frac{(4k^{1-1/q}+2k_q(x)^{1-1/q})B_q(k,x)^{1-1/q}\mu}{\beta_{q,B_q(k,x)}^2(A)}+(4+(k_q(x)/k)^{1-1/q})\lVert x-x^{k}\rVert_{2,1}. \label{DSrobustl1}
	\end{align}
\end{theorem}

\noindent\\
{\bf Proof.} We assume that $S$ is the block index set over the $k$ blocks with largest $\ell_2$-norms of $x$ such that $\lVert x_{S^c}\rVert_{2,1}=\lVert x-x^k\rVert_{2,1}$ and let $h=\hat{x}-x$. Recall that (\ref{ineq}) also holds here. Observe that \begin{align*}
\lVert x+h\rVert_{2,1}=\lVert x_S+h_S\rVert_{2,1}+\lVert x_{S^c}+h_{S^c}\rVert_{2,1}\geq \lVert x_S\rVert_{2,1}-\lVert h_S\rVert_{2,1}-\lVert x_{S^c}\rVert_{2,1}+\lVert h_{S^c}\rVert_{2,1},
\end{align*} and $\lVert x+h\rVert_{2,q}\leq \lVert x\rVert_{2,q}+\lVert h\rVert_{2,q}$, we can obtain that \begin{align*}
(\lVert x_S\rVert_{2,1}-\lVert h_S\rVert_{2,1}-\lVert x_{S^c}\rVert_{2,1}+\lVert h_{S^c}\rVert_{2,1})\cdot\lVert x\rVert_{2,q}\leq (\lVert x_S\rVert_{2,1}+\lVert x_{S^c}\rVert_{2,1})\cdot \lVert x\rVert_{2,q}+\lVert x\rVert_{2,1}\cdot\lVert h\rVert_{2,q}.
\end{align*}
Some simple manipulation yields \begin{align}
\lVert h_{S^c}\rVert_{2,1}\leq \lVert h_S\rVert_{2,1}+2\lVert x_{S^c}\rVert_{2,1}+\frac{\lVert x\rVert_{2,1}}{\lVert x\rVert_{2,q}}\lVert h\rVert_{2,q}=\lVert h_S\rVert_{2,1}+2\lVert x_{S^c}\rVert_{2,1}+k_q(x)^{1-1/q}\lVert h\rVert_{2,q},
\end{align}
which implies \begin{align}
\lVert h\rVert_{2,1}=\lVert h_S\rVert_{2,1}+\lVert h_{S^c}\rVert_{2,1}&\leq 2\lVert h_S\rVert_{2,1}+2\lVert x_{S^c}\rVert_{2,1}+k_q(x)^{1-1/q}\lVert h\rVert_{2,q}  \nonumber\\
&\leq (2k^{1-1/q}+k_q(x)^{1-1/q})\lVert h\rVert_{2,q}+2\lVert x_{S^c}\rVert_{2,1}. \label{errorbound}
\end{align}

(1) As for the problem (\ref{blocknorm_ratio}), suppose that $h\neq 0$ and $\lVert h\rVert_{2,q}>\frac{2\eta}{\beta_{q,B_q(k,x)}(A)}$, otherwise (\ref{robust}) holds trivially. Due to $\lVert Ah\rVert_2\leq 2\eta$, see (\ref{Ah_bound}), it follows that $\lVert h\rVert_{2,q}>\frac{\lVert Ah\rVert_2}{\beta_{q,B_q(k,x)}(A)}$. Then it yields \begin{align}
&\frac{\lVert Ah\rVert_2}{\lVert h\rVert_{2,q}}<{\beta_{q,B_q(k,x)}(A)}=\min\limits_{z\neq 0, k_q(z)\leq B_q(k,x)}\frac{\lVert Az\rVert_2}{\lVert z\rVert_{2,q}} \nonumber \\
&\quad \Rightarrow k_q(h)>B_q(k,x)\Rightarrow \lVert h\rVert_{2,1}>B_q(k,x)^{1-1/q}\lVert h\rVert_{2,q}=(4k^{1-1/q}+k_q(x)^{1-1/q})\lVert h\rVert_{2,q}.
\end{align}
Together with (\ref{errorbound}), we infer that $\lVert h\rVert_{2,q}<k^{1/q-1}\lVert x_{S^c}\rVert_{2,1}$, which completes the proof of (\ref{robust}). The error $\ell_{2,1}$ norm bound (\ref{robustl1}) follows immediately from (\ref{robust}) and (\ref{errorbound}). 

(2) In the case of the problem (\ref{DS}), we assume that $h\neq 0$ and $\lVert h\rVert_{2,q}>\frac{2B_q(k,x)^{1-1/q}\mu}{\beta_{q,B_q(k,x)}^2(A)}$, otherwise (\ref{DSrobust}) holds trivially. Since in this case $\lVert Ah\rVert_2^2\leq 2\mu\lVert h\rVert_{2,1}$, it follows that $\lVert h\rVert_{2,q}>\frac{B_q(k,x)^{1-1/q}}{\beta_{q,B_q(k,x)}^2(A)}\frac{\lVert Ah\rVert_2^2}{\lVert h\rVert_{2,1}}$. Then we get \begin{align}
&{\beta_{q,B_q(k,x)}^2(A)}=\min\limits_{z\neq 0, k_q(z)\leq B_q(k,x)}\frac{\lVert Az\rVert_2^2}{\lVert z\rVert_{2,q}^2}> \frac{\lVert Ah\rVert_2^2}{\lVert h\rVert_{2,q}^2}\cdot \left(\frac{B_q(k,x)}{k_q(h)}\right)^{1-1/q}\nonumber \\
&\quad \Rightarrow k_q(h)>B_q(k,x)\Rightarrow \lVert h\rVert_{2,1}>B_q(k,x)^{1-1/q}\lVert h\rVert_{2,q}=(4k^{1-1/q}+k_q(x)^{1-1/q})\lVert h\rVert_{2,q}.
\end{align}
Combining (\ref{errorbound}), we have $\lVert h\rVert_{2,q}<k^{1/q-1}\lVert x_{S^c}\rVert_{2,1}$, which completes the proof of (\ref{DSrobust}). The error $\ell_{2,1}$ norm bound (\ref{DSrobustl1}) follows immediately from (\ref{DSrobust}) and (\ref{errorbound}). The proof of Theorem 2 is now completed.

\subsection{Unconstrained Model}

To date, there has been no research carried out on the rigorous stable and robust analysis of the unconstrained model. For the first time this subsection of this paper seeks to investigate the theoretical recovery analysis results for the unconstrained version of (\ref{noiseless}), that is when $1<q\leq \infty$, we consider the problem \begin{align}
\min\limits_{z\in\mathbb{R}^N}\, \frac{1}{2}\lVert y-Az\rVert_2^2+\lambda \frac{\lVert z\rVert_{2,1}}{\lVert z\rVert_{2,q}}, \label{constrained_model}
\end{align}
where $y=Ax+\varepsilon$ and $\lambda>0$ is the regularization parameter. 

As has already been done for the constrained models, this subsection provides the recovery analysis results for the problem (\ref{constrained_model}) based on the $q$-ratio BCMSV. We start with the following main result for the case that the true signal $x$ is exactly block sparse.

\begin{theorem}
	If $x$ is block $k$-sparse, $q\in(1,\infty]$ and $\beta_{q,\left(\frac{3}{1-\kappa}\right)^{\frac{q}{q-1}}k}(A)>0$ with some  $\kappa\in(0,1)$, when\begin{align*} \lambda>\max\left(\frac{1-\kappa}{2(\kappa+2)}\lVert \varepsilon\rVert_2^2k^{1/q-1},\frac{\lVert A^{T}\varepsilon\rVert_{2,\infty}}{\kappa}\cdot\frac{\lVert y\rVert_2}{\beta_{q,\left(\frac{3}{1-\kappa}\right)^{\frac{q}{q-1}}k}(A)}\right),
	\end{align*} then the solution $\hat{x}$ to the problem (\ref{constrained_model}) satisfies \begin{align}
	\lVert \hat{x}-x\rVert_{2,q}&\leq \frac{k^{1-1/q}\lVert A\rVert_2}{\beta_{q,\left(\frac{3}{1-\kappa}\right)^{\frac{q}{q-1}}k}^2(A)}\cdot  \frac{3(\kappa+1)}{1-\kappa}\sqrt{\lVert \varepsilon\rVert_2^2+2k^{1-1/q}\lambda}, \\
	\lVert \hat{x}-x\rVert_{2,1}&\leq \frac{k^{2-2/q}\lVert A\rVert_2}{\beta_{q,\left(\frac{3}{1-\kappa}\right)^{\frac{q}{q-1}}k}^2(A)}\cdot  \frac{9(\kappa+1)}{(1-\kappa)^2}\sqrt{\lVert \varepsilon\rVert_2^2+2k^{1-1/q}\lambda}.
	\end{align}
\end{theorem}

\noindent\\
{\bf Proof.} Let $\alpha=z/\lVert z\rVert_{2,q}$ and $\beta=\lVert z\rVert_{2,q}$, then the problem (\ref{constrained_model}) is equivalent to solve \begin{align}
\min\limits_{\{\alpha\in\mathbb{R}^N,\beta\in\mathbb{R}^{+}\}} \frac{1}{2}\lVert y-\beta A\alpha\rVert_2^2+\lambda \lVert \alpha\rVert_{2,1}.\label{constained_modelNew}
\end{align}
If the solution to (\ref{constained_modelNew}) is $(\hat{\alpha},\hat{\beta})$, then the solution to (\ref{constrained_model}) is $\hat{x}=\hat{\beta}\hat{\alpha}$. By using the Karush-Kuhn-Tucker (KKT) condition of (\ref{constained_modelNew}) with respective to $\beta$, we have \begin{align*}
(A\hat{\alpha})^{T}(\hat{\beta}A\hat{\alpha}-y)=0,
\end{align*}
such that $\lVert \hat{x}\rVert_{2,q}=\hat{\beta}=\frac{(A\hat{\alpha})^T y}{\lVert A\hat{\alpha}\rVert_2^2}$. Therefore, we obtain that \begin{align*}
\lVert \hat{x}\rVert_{2,q}=\frac{(A\hat{\alpha})^T y}{\lVert A\hat{\alpha}\rVert_2^2}\leq \frac{\lVert A\hat{\alpha}\rVert_2\lVert y\rVert_2}{\lVert A\hat{\alpha}\rVert_2^2}\leq \frac{\lVert y\rVert_2}{\lVert A\hat{\alpha}\rVert_2}
\end{align*}
Since $\hat{x}$ is the solution to (\ref{constrained_model}), we have \begin{align*}
\frac{1}{2}\lVert y-A\hat{x}\rVert_2^2+\lambda \frac{\lVert \hat{x}\rVert_{2,1}}{\lVert \hat{x}\rVert_{2,q}}\leq 
\frac{\lVert \varepsilon\rVert_2^2}{2}+\lambda\frac{\lVert x\rVert_{2,1}}{\lVert x\rVert_{2,q}}\leq\frac{\lVert \varepsilon\rVert_2^2}{2}+\lambda k_{q}(x)^{1-1/q}\leq \frac{\lVert \varepsilon\rVert_2^2}{2}+\lambda k^{1-1/q},
\end{align*}
which implies that $\frac{\lVert \hat{x}\rVert_{2,1}}{\lVert \hat{x}\rVert_{2,q}}\leq \lVert \varepsilon\rVert_2^2/(2\lambda)+k^{1-1/q}$. Here we use the fact that $k_q(x)\leq k_0(x)=\lVert x\rVert_{2,0}\leq k$. Hence, when $\lambda\geq\frac{1-\kappa}{2(\kappa+2)}\lVert \varepsilon\rVert_2^2k^{1/q-1}$ with some $\kappa\in(0,1)$, we have $\frac{\lVert \hat{x}\rVert_{2,1}}{\lVert \hat{x}\rVert_{2,q}}\leq \left(\frac{\kappa+2}{1-\kappa}\right)k^{1-1/q}+k^{1-1/q}\leq \left(\frac{3}{1-\kappa}\right)k^{1-1/q}$. As a consequence of $\hat{x}=\hat{\beta}\hat{\alpha}$, it follows that $k_q(\hat{\alpha})\leq \left(\frac{3}{1-\kappa}\right)^{\frac{q}{q-1}}k$, which yields $\lVert A\hat{\alpha}\rVert_2\geq \beta_{q,\left(\frac{3}{1-\kappa}\right)^{\frac{q}{q-1}}k}(A)$ since $\lVert\hat{\alpha}\rVert_{2,q}=1$. To summarize what we have proved, we get \begin{align}
\lVert \hat{x}\rVert_{2,q}\leq \frac{\lVert y\rVert_2}{\beta_{q,\left(\frac{3}{1-\kappa}\right)^{\frac{q}{q-1}}k}(A)},
\end{align}
when $\lambda\geq\frac{1-\kappa}{2(\kappa+2)}\lVert \varepsilon\rVert_2^2k^{1/q-1}$ with $\kappa\in(0,1)$.

In addition, when $\lambda\geq \frac{\lVert A^{T}\varepsilon\rVert_{2,\infty}}{\kappa}\cdot\frac{\lVert y\rVert_2}{\beta_{q,\left(\frac{3}{1-\kappa}\right)^{\frac{q}{q-1}}k}(A)}\geq \frac{\lVert A^{T}\varepsilon\rVert_{2,\infty}}{\kappa}\lVert \hat{x}\rVert_{2,q}$, let $h=\hat{x}-x$, then it holds that \begin{align*}
\lambda \frac{\lVert \hat{x}\rVert_{2,1}}{\lVert \hat{x}\rVert_{2,q}}&\leq \frac{\lVert \varepsilon\rVert_2^2}{2}-\frac{1}{2}\lVert A(\hat{x}-x)-\varepsilon\rVert_2^2+\lambda\frac{\lVert x\rVert_{2,1}}{\lVert x\rVert_{2,q}} \\
&= \langle A(\hat{x}-x),\varepsilon\rangle-\lVert A(\hat{x}-x)\rVert_2^2+\lambda\frac{\lVert x\rVert_{2,1}}{\lVert x\rVert_{2,q}} \\
&\leq \langle Ah,\varepsilon\rangle+\lambda\frac{\lVert x\rVert_{2,1}}{\lVert x\rVert_{2,q}} \\
&\leq \lVert h\rVert_{2,1}\lVert A^{T}\varepsilon\rVert_{2,\infty}+\lambda\frac{\lVert x\rVert_{2,1}}{\lVert x\rVert_{2,q}} \\
&\leq \frac{\lambda\kappa\lVert h\rVert_{2,1}}{\lVert \hat{x}\rVert_{2,q}}+\lambda\frac{\lVert x\rVert_{2,1}}{\lVert x\rVert_{2,q}}.
\end{align*}
Then, we can obtain that \begin{align*}
\lVert x\rVert_{2,1}-\lVert h_S\rVert_{2,1}+\lVert h_{S^c}\rVert_{2,1}&\leq \lVert x+h\rVert_{2,1}\leq\kappa \lVert h\rVert_{2,1}+\frac{\lVert x\rVert_{2,1}}{\lVert x\rVert_{2,q}}\cdot\lVert x+h\rVert_{2,q} \\
&\leq \kappa \lVert h\rVert_{2,1}+\lVert x\rVert_{2,1}+\frac{\lVert x\rVert_{2,1}}{\lVert x\rVert_{2,q}}\lVert h\rVert_{2,q}\\
&\leq \kappa \lVert h\rVert_{2,1}+\lVert x\rVert_{2,1}+k^{1-1/q}\lVert h\rVert_{2,q},
\end{align*}
which implies that $\lVert h_{S^c}\rVert_{2,1}\leq \lVert h_S\rVert_{2,1}+\kappa\lVert h\rVert_{2,1}+k^{1-1/q}\lVert h\rVert_{2,q}$. As a result, \begin{align*}
\lVert h\rVert_{2,1}\leq \lVert h_S\rVert_{2,1}+\lVert h_{S^c}\rVert_{2,1}&\leq 2\lVert h_{S}\rVert_{2,1}+\kappa\lVert h\rVert_{2,1}+k^{1-1/q}\lVert h\rVert_{2,q}  \\
&\leq 3k^{1-1/q}\lVert h\rVert_{2,q}+\kappa\lVert h\rVert_{2,1}.
\end{align*}
Therefore, it follows that $(1-\kappa)\lVert h\rVert_{2,1}\leq 3k^{1-1/q}\lVert h\rVert_{2,q}$, i.e., $k_q(h)=(\lVert h\rVert_{2,1}/\lVert h\rVert_{2,q})^{\frac{q}{q-1}}\leq \left(\frac{3}{1-\kappa}\right)^{\frac{q}{q-1}}k$. 

Moreover, $\lVert Ah\rVert_2^2=\langle Ah,Ah \rangle\leq \lVert h\rVert_{2,1}\lVert A^{T}Ah\rVert_{2,\infty}$, and
\begin{align*}
\lVert A^{T}Ah\rVert_{2,\infty}&\leq \lVert A^{T}(y-Ax)\rVert_{2,\infty}+\lVert A^{T}(y-A\hat{x})\rVert_{2,\infty} \\
&\leq \lVert A^{T}\varepsilon\rVert_{2,\infty}+\lVert A^{T}(y-A\hat{x})\rVert_{2,\infty}.
\end{align*}
Meanwhile, the KKT condition of (\ref{constained_modelNew}) with respective to $\alpha$ implies that \begin{align}
\hat{\beta}A^{T}(A\hat{x}-y)=-\lambda\partial \lVert \hat{\alpha}\rVert_{2,1}, \label{KKT}
\end{align}
where the sub-gradients in $\partial \lVert \hat{\alpha}\rVert_{2,1}$ for the $i$-th block are
$\hat{\alpha}[i]/\lVert \hat{\alpha}[i]\rVert_2$ when $\hat{\alpha}[i]\neq 0$ and is some vector $g$ satisfying $\lVert g\rVert_2\leq 1$ when $\hat{\alpha}[i]=0$. Therefore, we get $\hat{\beta}\lVert A^{T}(A\hat{x}-y)\rVert_{2,\infty}\leq \lambda$, i.e., $\lVert  A^{T}(A\hat{x}-y)\rVert_{2,\infty}\leq \frac{\lambda}{\lVert \hat{x}\rVert_{2,q}}$. As a consequence of $\lambda\geq \frac{\lVert A^{T}\varepsilon\rVert_{2,\infty}}{\kappa}\lVert \hat{x}\rVert_{2,q}$, it holds that \begin{align*}
\lVert A^{T}Ah\rVert_{2,\infty}\leq \lVert A^{T}\varepsilon\rVert_{2,\infty}+\lVert A^{T}(y-A\hat{x})\rVert_{2,\infty} \leq (\kappa+1)\frac{\lambda}{\lVert \hat{x}\rVert_{2,q}},
\end{align*}
which leads to $\lVert Ah\rVert_2^2\leq \lVert h\rVert_{2,1}\lVert A^{T}Ah\rVert_{2,\infty}\leq (\kappa+1)\frac{\lambda}{\lVert \hat{x}\rVert_{2,q}}\lVert h\rVert_{2,1}$.

Then, with $k_q(h)\leq \left(\frac{3}{1-\kappa}\right)^{\frac{q}{q-1}}k$, \begin{align*}
\beta_{q,\left(\frac{3}{1-\kappa}\right)^{\frac{q}{q-1}}k}^2(A)\lVert h\rVert_{2,q}^2\leq \lVert Ah\rVert_2^2&\leq (\kappa+1)\frac{\lambda}{\lVert \hat{x}\rVert_{2,q}}\lVert h\rVert_{2,1}\\
&\leq \frac{3(\kappa+1)}{1-\kappa}k^{1-1/q}\lVert h\rVert_{2,q}\cdot\frac{\lambda}{\lVert \hat{x}\rVert_{2,q}},
\end{align*}
which implies that \begin{align*}
\lVert h\rVert_{2,q}\leq \frac{k^{1-1/q}}{\beta_{q,\left(\frac{3}{1-\kappa}\right)^{\frac{q}{q-1}}k}^2(A)}\cdot  \frac{3(\kappa+1)}{1-\kappa}\cdot\frac{\lambda}{\lVert \hat{x}\rVert_{2,q}}.
\end{align*}

Finally, we establish an upper bound for $\frac{\lambda}{\lVert \hat{x}\rVert_{2,q}}$. By using the KKT condition (\ref{KKT}) again, we can obtain that \begin{align*}
\hat{\beta}\lVert A^{T}(A\hat{x}-y)\rVert_2\geq \lambda\sqrt{\lVert \hat{\alpha}\rVert_{2,0}}\geq \lambda.
\end{align*}
Hence, it follows that \begin{align*}
\frac{\lambda}{\lVert \hat{x}\rVert_{2,q}}\leq \lVert A^{T}(A\hat{x}-y)\rVert_2\leq \lVert A^{T}\rVert_2\lVert A\hat{x}-y\rVert_2\leq \lVert A\rVert_2\sqrt{\lVert \varepsilon\rVert_2^2+2k^{1-1/q}\lambda}.
\end{align*}
Therefore, we get \begin{align*}
\lVert h\rVert_{2,q}\leq\frac{k^{1-1/q}\lVert A\rVert_2}{\beta_{q,\left(\frac{3}{1-\kappa}\right)^{\frac{q}{q-1}}k}^2(A)}\cdot  \frac{3(\kappa+1)}{1-\kappa}\sqrt{\lVert \varepsilon\rVert_2^2+2k^{1-1/q}\lambda},
\end{align*}
and \begin{align*}
\lVert h\rVert_{2,1}\leq \left(\frac{3}{1-\kappa}\right)k^{1-1/q}\lVert h\rVert_{2,q}\leq \frac{k^{2-2/q}\lVert A\rVert_2}{\beta_{q,\left(\frac{3}{1-\kappa}\right)^{\frac{q}{q-1}}k}^2(A)}\cdot  \frac{9(\kappa+1)}{(1-\kappa)^2}\sqrt{\lVert \varepsilon\rVert_2^2+2k^{1-1/q}\lambda}.
\end{align*}
The proof is now completed.\\

Furthermore, the corresponding result for the case that the true signal is not exactly block sparse can be obtained as follows. 

\begin{theorem}
	For any $x\in\mathbb{R}^N$, $q\in (1,\infty]$ and some $\kappa\in(0,1)$, we denote $B_{\kappa,q}(k,x)=\left(\frac{4k^{1-1/q}}{1-\kappa}+\frac{k_q(x)^{1-1/q}}{1-\kappa}\right)^{\frac{q}{q-1}}$. If $\lambda\geq \max\left(\frac{1-\kappa}{2(\kappa+2)}\lVert \varepsilon\rVert_2^2k_q(x)^{1/q-1},\frac{\lVert A^{T}\varepsilon\rVert_{2,\infty}}{\kappa}\cdot\frac{\lVert y\rVert_2}{\beta_{q,\left(\frac{3}{1-\kappa}\right)^{\frac{q}{q-1}}k_q(x)}(A)}\right)$, then the solution $\hat{x}$ to (\ref{constrained_model}) obeys \begin{align}
	\lVert \hat{x}-x\rVert_{2,q}&\leq \frac{(\kappa+1)\lVert A\rVert_2}{\beta_{q,\frac{B_{\kappa,q}(k,x)}{(1-\kappa)^{\frac{q}{q-1}}}}^2(A)}\cdot  \frac{B_{\kappa,q}(k,x)^{1-1/q}}{1-\kappa}\cdot\sqrt{\lVert \varepsilon\rVert_2^2+2k_q(x)^{1-1/q}\lambda}+k^{1/q-1}\lVert x-x^{k}\rVert_{2,1}, \label{constrained_robust} \\
	\lVert\hat{x}-x\rVert_{2,1}&\leq \frac{(\kappa+1)(\frac{2k^{1-1/q}}{1-\kappa}+\frac{k_q(x)^{1-1/q}}{1-\kappa})\lVert A\rVert_2}{\beta_{q,\frac{B_{\kappa,q}(k,x)}{(1-\kappa)^{\frac{q}{q-1}}}}^2(A)}\cdot  \frac{B_{\kappa,q}(k,x)^{1-1/q}}{1-\kappa}\cdot\sqrt{\lVert \varepsilon\rVert_2^2+2k_q(x)^{1-1/q}\lambda} \nonumber \\
	&\quad +\left(\frac{4}{1-\kappa}+\frac{(k_q(x)/k)^{1-1/q}}{1-\kappa}\right)\lVert x-x^{k}\rVert_{2,1}, \label{constrained_robust1}
	\end{align}
	\end{theorem}

\noindent\\
{\bf Proof.}  We let $h=\hat{x}-x$ and suppose $S$ is the block index set over the $k$ blocks with largest $\ell_2$-norms of $x$ such that $\lVert x_{S^c}\rVert_{2,1}=\lVert x-x^k\rVert_{2,1}$. Following similar arguments in the Proof of Theorem 3, when $\lambda\geq \max\left(\frac{1-\kappa}{2(\kappa+2)}\lVert \varepsilon\rVert_2^2k_q(x)^{1/q-1},\frac{\lVert A^{T}\varepsilon\rVert_{2,\infty}}{\kappa}\cdot\frac{\lVert y\rVert_2}{\beta_{q,\left(\frac{3}{1-\kappa}\right)^{\frac{q}{q-1}}k_q(x)}(A)}\right)$, we have \begin{align*}
\lVert \hat{x}\rVert_{2,1}&\leq \kappa\lVert h\rVert_{2,1}+\frac{\lVert x\rVert_{2,1}}{\lVert x\rVert_{2,q}}\cdot\lVert x+h\rVert_{2,q} \\
&\leq \kappa\lVert h\rVert_{2,1}+\lVert x\rVert_{2,1}+k_q(x)^{1-1/q}\lVert h\rVert_{2,q}.
\end{align*}
Hence, \begin{align*}
\lVert x\rVert_{2,1}=\lVert x_S\rVert_{2,1}+\lVert x_{S^c}\rVert_{2,1}&\geq \lVert \hat{x}\rVert_{2,1}-\kappa\lVert h\rVert_{2,1}-k_{q}(x)^{1-1/q}\lVert h\rVert_{2,q} \\
&=\lVert x_S+x_{S^c}+h_S+h_{S^c}\rVert_{2,1}-\kappa\lVert h_S+h_{S^c}\rVert_{2,1}-k_{q}(x)^{1-1/q}\lVert h\rVert_{2,q} \\
&\geq \lVert x_S\rVert_{2,1}-\lVert x_{S^c}\rVert_{2,1}+(1-\kappa)\lVert h_{S^c}\rVert_{2,1}-(1+\kappa)\lVert h_{S}\rVert_{2,1}-k_{q}(x)^{1-1/q}\lVert h\rVert_{2,q},
\end{align*}
which implies that \begin{align*}
\lVert h_{S^c}\rVert_{2,1}\leq \frac{1+\kappa}{1-\kappa}\lVert h_S\rVert_{2,1}+\frac{2}{1-\kappa}\lVert x_{S^c}\rVert_{2,1}+\frac{k_q(x)^{1-1/q}}{1-\kappa}\lVert h\rVert_{2,q}.
\end{align*}
As a result, \begin{align}
\lVert h\rVert_{2,1}=\lVert h_S\rVert_{2,1}+\lVert h_{S^c}\rVert_{2,1}&\leq \frac{2}{1-\kappa}\lVert h_S\rVert_{2,1}+\frac{2}{1-\kappa}\lVert x_{S^c}\rVert_{2,1}+\frac{k_q(x)^{1-1/q}}{1-\kappa}\lVert h\rVert_{2,q} \nonumber \\
&\leq \left(\frac{2k^{1-1/q}}{1-\kappa}+\frac{k_q(x)^{1-1/q}}{1-\kappa}\right)\lVert h\rVert_{2,q}+\frac{2}{1-\kappa}\lVert x_{S^c}\rVert_{2,1}. \label{constrained_errorbound}
\end{align}

We assume that $h\neq 0$ and $\lVert h\rVert_{2,q}> \frac{(\kappa+1)\lVert A\rVert_2}{\beta_{q,\frac{B_{\kappa,q}(k,x)}{(1-\kappa)^{\frac{q}{q-1}}}}^2(A)}\cdot  \frac{B_{\kappa,q}(k,x)^{1-1/q}}{1-\kappa}\cdot\sqrt{\lVert \varepsilon\rVert_2^2+2k_q(x)^{1-1/q}\lambda}$, otherwise (\ref{constrained_robust}) holds trivially. Because $\lVert Ah\rVert_2^2\leq (\kappa+1)\frac{\lambda}{\lVert \hat{x}\rVert_{2,q}}\lVert h\rVert_{2,1}\leq (\kappa+1)\lVert A\rVert_2\sqrt{\lVert \varepsilon\rVert_2^2+2k_q(x)^{1-1/q}\lambda}\lVert h\rVert_{2,1} $, so we have $\lVert h\rVert_{2,q}>\frac{B_{\kappa,q}(k,x)^{1-1/q}/(1-\kappa)}{\beta_{q,\frac{B_{\kappa,q}(k,x)}{(1-\kappa)^{\frac{q}{q-1}}}}^2(A)}\frac{\lVert Ah\rVert_2^2}{\lVert h\rVert_{2,1}}$. Then we get \begin{align}
&{\beta_{q,\frac{B_{\kappa,q}(k,x)}{(1-\kappa)^{\frac{q}{q-1}}}}^2(A)}=\min\limits_{z\neq 0, k_q(z)\leq \frac{B_{\kappa,q}(k,x)}{(1-\kappa)^{\frac{q}{q-1}}}}\frac{\lVert Az\rVert_2^2}{\lVert z\rVert_{2,q}^2}> \frac{\lVert Ah\rVert_2^2}{\lVert h\rVert_{2,q}^2}\cdot \left(\frac{B_{\kappa,q}(k,x)}{k_q(h)(1-\kappa)^{\frac{q}{q-1}}}\right)^{1-1/q}\nonumber \\
&\quad \Rightarrow k_q(h)>\frac{B_{\kappa,q}(k,x)}{(1-\kappa)^{\frac{q}{q-1}}}\Rightarrow \lVert h\rVert_{2,1}>\frac{B_{\kappa,q}(k,x)^{1-1/q}}{1-\kappa}\lVert h\rVert_{2,q}=\left(\frac{4}{1-\kappa}k^{1-1/q}+\frac{k_q(x)^{1-1/q}}{1-\kappa}\right)\lVert h\rVert_{2,q}.
\end{align}
Combining (\ref{constrained_errorbound}), we have $\lVert h\rVert_{2,q}<k^{1/q-1}\lVert x_{S^c}\rVert_{2,1}$, which completes the proof of (\ref{constrained_robust}). The error $\ell_{2,1}$ norm bound (\ref{constrained_robust1}) follows immediately from (\ref{constrained_robust}) and (\ref{constrained_errorbound}). \\

At the end of this subsection, it should be pointed out the algorithms proposed in \cite{huang2018sparse} for the generalized entropy function minimization problem can be used here in solving the unconstrained model (\ref{constrained_model}) with some careful generalizations from non-block to block setting. Further work is required to evaluate the performances of the algorithms, which is out of the scope of this paper. Instead, in the following section we provide algorithms via a convex-concave procedure to solve the constrained models (\ref{blocknorm_ratio}) and (\ref{DS}).

\section{Algorithms}

In fact, the minimization problems (\ref{blocknorm_ratio}) and (\ref{DS}) belong to the nonlinear fractional programming, where both the numerator and denominator are convex functions. This specific nonlinear fractional programming was comprehensively discussed in Chapter 4 of \cite{stancu2012fractional}, see also \cite{schaible1976minimization,schaible2004recent}. Basically there are two kinds of methods to solve it, namely parametric methods and a change of variable method. \begin{itemize}
	\item \underline{Parametric methods}.
	To solve the fractional problems (\ref{blocknorm_ratio}) and (\ref{DS}), a class of methods by iteratively solving the following difference of convex functions problems depending on a parameter $\lambda\in\mathbb{R}$: 
	\begin{subnumcases}{\min\limits_{z\in\mathbb{R}^N}  \lambda\lVert z\rVert_{2,1}-\lVert z\rVert_{2,q},\quad \text{subject to} \quad}
	\lVert y-Az\rVert_2\leq \eta, \\
	\lVert A^{T}(y-Az)\rVert_{2,\infty}\leq \mu.
	\end{subnumcases}
    can be used, see \cite{zhou2020minimization} for detailed arguments within a non-block framework. However, how to solve these subproblems efficiently are left for future work.
	\item \underline{Change of variable method}. As a more direct and faster solver compared to the parametric methods, in this section we mainly focus on the block version of the change of variable method proposed in \cite{zhou2020minimization} and provide detailed discussions in what follows.
\end{itemize}

With a change of variable by letting $v=\frac{z}{\lVert z\rVert_{2,1}}\in\mathbb{R}^N$ and $t=\frac{1}{\lVert z\rVert_{2,1}}\in\mathbb{R^{+}}$, the minimization problems (\ref{blocknorm_ratio}) and (\ref{DS}) are equivalent to the following problems
\begin{subnumcases}{\min\limits_{v\in\mathbb{R}^N, t\in\mathbb{R}^{+}} \frac{1}{\lVert v\rVert_{2,q}} \quad \text{subject to} \quad}
t>0, \lVert y-Av/t\rVert_2\leq \eta\,\, \text{and}\,\,\lVert v\rVert_{2,1}=1, \label{cvm} \\
t>0, \lVert A^{T}(y-Av/t)\rVert_{2,\infty}\leq \mu \,\, \text{and}\,\, \lVert v\rVert_{2,1}=1. \label{cvm_DS}
\end{subnumcases}

Then we are able to replace the equality constraint $\lVert v\rVert_{2,1}=1$ by $\lVert v\rVert_{2,1}\leq 1$, and change the minimization problem to a maximization problem, please see the arguments in Section 3 of \cite{schaible1976minimization} for details. Henceforth, it suffices to solve
\begin{subnumcases}{\max\limits_{v\in\mathbb{R}^N, t\in\mathbb{R}^{+}} \lVert v\rVert_{2,q}\quad \text{subject to} \quad}
t\geq t_0, \lVert y-Av/t\rVert_2\leq \eta\,\, \text{and}\,\,\lVert v\rVert_{2,1}\leq 1, \label{maxq} \\
t\geq t_0, \lVert A^{T}(y-Av/t)\rVert_{2,\infty}\leq \mu \,\, \text{and}\,\, \lVert v\rVert_{2,1}\leq 1, \label{maxq_DS}
\end{subnumcases}
where $t_0=1/a$ for some $a\geq \max\{\lVert z\rVert_{2,1}| \lVert y-Az\rVert_2\leq \eta\,\,\mathrm{or}\,\, \lVert A^{T}(y-Az)\rVert_{2,\infty}\leq \mu \}$. If its solution is denoted as $\hat{v}$ and $\hat{t}$, then our final recovered signal goes to $\hat{x}=\hat{v}/\hat{t}$.

Hereafter, an algorithm via a convex-concave procedure (CCP) \cite{lipp2016variations} is adopted to solve this convex-concave problem. The corresponding CCP algorithm goes as follows: 

\begin{algorithm}	
	\caption{CCP to solve (\ref{maxq}) and (\ref{maxq_DS}).}
	\begin{algorithmic}
		\STATE {Input: measurement matrix $A$, measurement vector $y$, error bounds $\eta$ or $\mu$, and $t_0$.}\\
		\STATE {Initialization: Given an initial point $v^{0}=\frac{x^{(0)}}{\lVert x^{(0)}\rVert_{2,1}}$ and $k=0$.}\\
		\STATE {Iteration: Repeat until a stopping criterion is met at $k=\bar{n}$. }\\
		\STATE {1. Convexify: Linearize $\lVert v\rVert_{2,q}$ with the approximation $$\lVert v^{(k)}\rVert_{2,q}+\nabla (\lVert v\rVert_{2,q})_{v=v^{(k)}}^T(v-v^{(k)})=\lVert v^{(k)}\rVert_{2,q}+\left[\lVert v^{(k)}\rVert_{2,q}^{1-q}v_{*}^{(k)}\odot v^{(k)}\right]^T(v-v^{(k)}),$$
		where $v_{*}^{(k)}=\left[\underbrace{\lVert v^{(k)}[1]\rVert_2^{q-2}, \cdots,\lVert v^{(k)}[1]\rVert_2^{q-2}}_{d},\underbrace{\lVert v^{(k)}[2]\rVert_2^{q-2},\cdots,\lVert v^{(k)}[2]\rVert_2^{q-2}}_{d},\cdots,\underbrace{\lVert v^{(k)}[M]\rVert_2^{q-2},\cdots,\lVert v^{(k)}[M]\rVert_2^{q-2}}_{d}\right]^T$ with $\lVert v^{(k)}[i]\rVert_2$ denoting the $\ell_2$ norm of the $i$-th block of $v^{(k)}$ for $i\in[M]$, and $\odot$ denoting the Hadamard product.
}
		\STATE {2. Solve: Set the value of $v^{(k+1)}\in\mathbb{R}^N, t^{(k+1)}\in \mathbb{R}^{+}$ to be a solution of \begin{align}
			&\max\limits_{v\in\mathbb{R}^N, t\in\mathbb{R}^{+}}\,\lVert v^{(k)}\rVert_{2,q}+\left[\lVert v^{(k)}\rVert_{2,q}^{1-q}v_{*}^{(k)}\odot v^{(k)}\right]^T(v-v^{(k)})  \nonumber \\
			&\quad\text{s.t. \, $t\geq t_0$, $\lVert v\rVert_{2,1}\leq 1$, $\lVert y-Av/t\rVert_2\leq \eta$ for (\ref{maxq}) \Big\{$\lVert A^{T}(y-Av/t)\rVert_{2,\infty}\leq \mu$  for (\ref{maxq_DS})\Big\}   }. \label{ccp} \end{align} }
		
		\vspace{-1.5em}
		\STATE {3. Update iteration: $k=k+1$.} \\
		\STATE {Output: The recovered signal $\hat{x}=v^{(\bar{n})}/{t^{(\bar{n})}}$.}
	\end{algorithmic}
\end{algorithm}

In this algorithm, we use the solution of the mixed $\ell_2/\ell_1$-minimization problem (\ref{mixedl2l1}) as our $x^{(0)}$. In the case of $q=\infty$,  the linearized term for $\lVert v\Vert_{2,\infty}$ at $v^{(k)}$ will be $\lVert v^{(k)}\rVert_{2,\infty}+\left(\frac{v^{(k)}[j]}{\lVert v^{(k)}[j]\rVert_2}\right)^T(v[j]-v^{(k)}[j])$  if the block index to achieve the $\ell_{2,\infty}$ norm of $v^{(k)}$ is $j$, i.e., $\lVert v^{(k)}[j]\rVert_2=\lVert v^{(k)}\rVert_{2,\infty}$. We solve all the convex sub-programs such as (\ref{ccp}) using the CVX toolbox in Matlab \cite{grant2014cvx}.

\section{Numerical experiments}

In this section, we conduct numerical experiments to illustrate the performance of our proposed method in block sparse signal reconstruction from different perspectives. In all the following experiments, the block $k$-sparse signal is generated by choosing $k$ blocks uniformly at random, and then choosing the non-zero values from the standard normal distribution for these $k$ blocks. 

\subsection{A test}
In this set of experiments, we tested the CCP recovery algorithms for a block sparse signal $x\in\mathbb{R}^{400}$ reconstruction with a Gaussian random measurement matrix $A\in\mathbb{R}^{120\times 400}$. We fixed the block size $d=2$. We considered two cases, one is that the true signal $x$ has a block sparsity level of $20$ and the measurements are noise free, the other case is that the true signal is block $10$-sparse and the measurements are noisy with either $\ell_2$-bounded noise or $\ell_{2,\infty}$-bounded noise. For the $\ell_2$-bounded noise, we set $\varepsilon=0.1u/\lVert u\rVert_2$ to ensure that $\lVert \varepsilon\rVert_2\leq 0.1$ with $u=\mathrm{randn}(120,1)$, while for the $\ell_{2,\infty}$-bounded noise, we set $\varepsilon=0.1u/\lVert A^{T}u\rVert_{2,\infty}$ such that $\lVert A^{T}\varepsilon\rVert_{2,\infty}\leq 0.1$.

\begin{figure}[htbp]
	\centering
	\includegraphics[width=\textwidth,height=0.35\textheight]{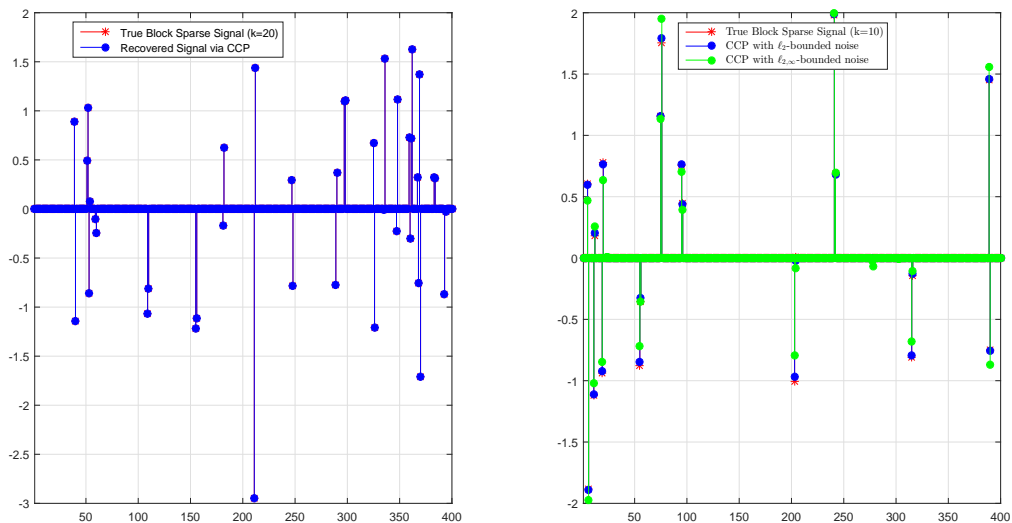} 
	\caption{A numerical test for the $\ell_{2,1}/\ell_{2,q}$ minimization via CCP with $q=2$, left panel (Noise Free) and right panel (Noisy with $\eta=\mu=0.1$).} \label{CCP_Test}
\end{figure}

As shown in Figure \ref{CCP_Test}, when there is no noise in the measurements, a perfect recovery can be achieved via the $\ell_{2,1}/\ell_{2,q}$ minimization problem with $q=2$, while it would not be impaired almost at all with slightly noisy measurements regardless the noise types.  

\subsection{Different choices of $q$}

In this subsection, we compared the performances of the proposed $\ell_{2,1}/\ell_{2,q}$ minimization problem with different $q$ varying from $1.1,1.5,2,4,\infty$. In this study, $A$ is an $m\times 200$ random matrix generated as Gaussian with $m=\{20,40,60,80,100,120,140\}$. The true signal $x\in\mathbb{R}^{200}$ is simulated as block $10$-sparse with block size $d=2$. For each $q$, we replicated the noiseless experiments 100 times with different $A$ and $x$. It is recorded as one success if the relative error $\frac{\lVert \hat{x}-x\rVert_2}{\lVert x\rVert_2}\leq 10^{-3}$. 

\begin{figure}[htbp]
	\centering
	\includegraphics[width=\textwidth,height=0.4\textheight]{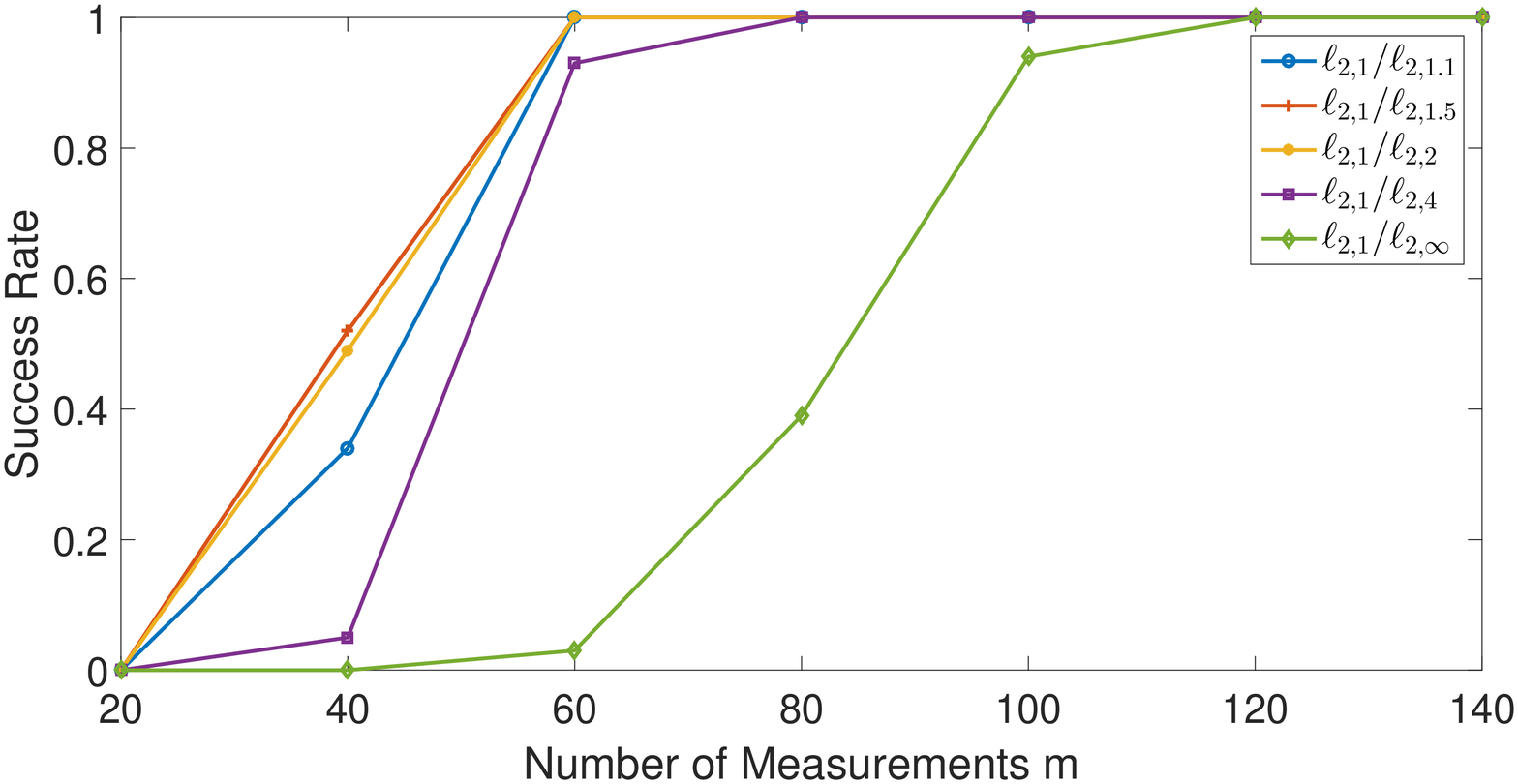} 
	\caption{Reconstruction performance comparison for the $\ell_{2,1}/\ell_{2,q}$ minimization with  Gaussian random measurements and different $q$ varying from $1.1,1.5,2,4,\infty$.} \label{differentq}
\end{figure}

Figure \ref{differentq} shows the success rate over the 100 replicates for various values of parameter $q$ and number of measurements $m$. It can be seen that $q=1.5$ is the best among all tested values of $q$, and the results for $q=1.1$ and $q=2$ are better than those for $q=4$ and $q=\infty$. The ability to choose suitable values of $q$ enables us to fully exploit the block sparsity promoting power of the proposed models and to achieve better reconstruction performances.

\subsection{Comparison on different block sparse recovery methods}

In the final part of this section, comparisons between the proposed $\ell_{2,1}/\ell_{2,1.5}$ minimization and other state-of-the-art block sparse signal recovery methods including mixed $\ell_2/\ell_{p}$ with $p=0.2,0.5,0.8$, group lasso and $\ell_2/\ell_{1-2}$ are performed. For each recovery method, we replicated the noiseless experiments $100$ times with different $A$ and $x$ and evaluated its performance in terms of success rate.

\subsubsection{Gaussian random matrices}
We start with the comparison of block sparse signal recovery by using the Gaussian random matrices as the measurement matrices. We set $m=60, N=200$ and choose the block size $d=2$, and the block sparsity level $k=\{7,9,11,13,15,17,19,21,23\}$. 

\begin{figure}[htbp]
	\centering
	\includegraphics[width=\textwidth,height=0.4\textheight]{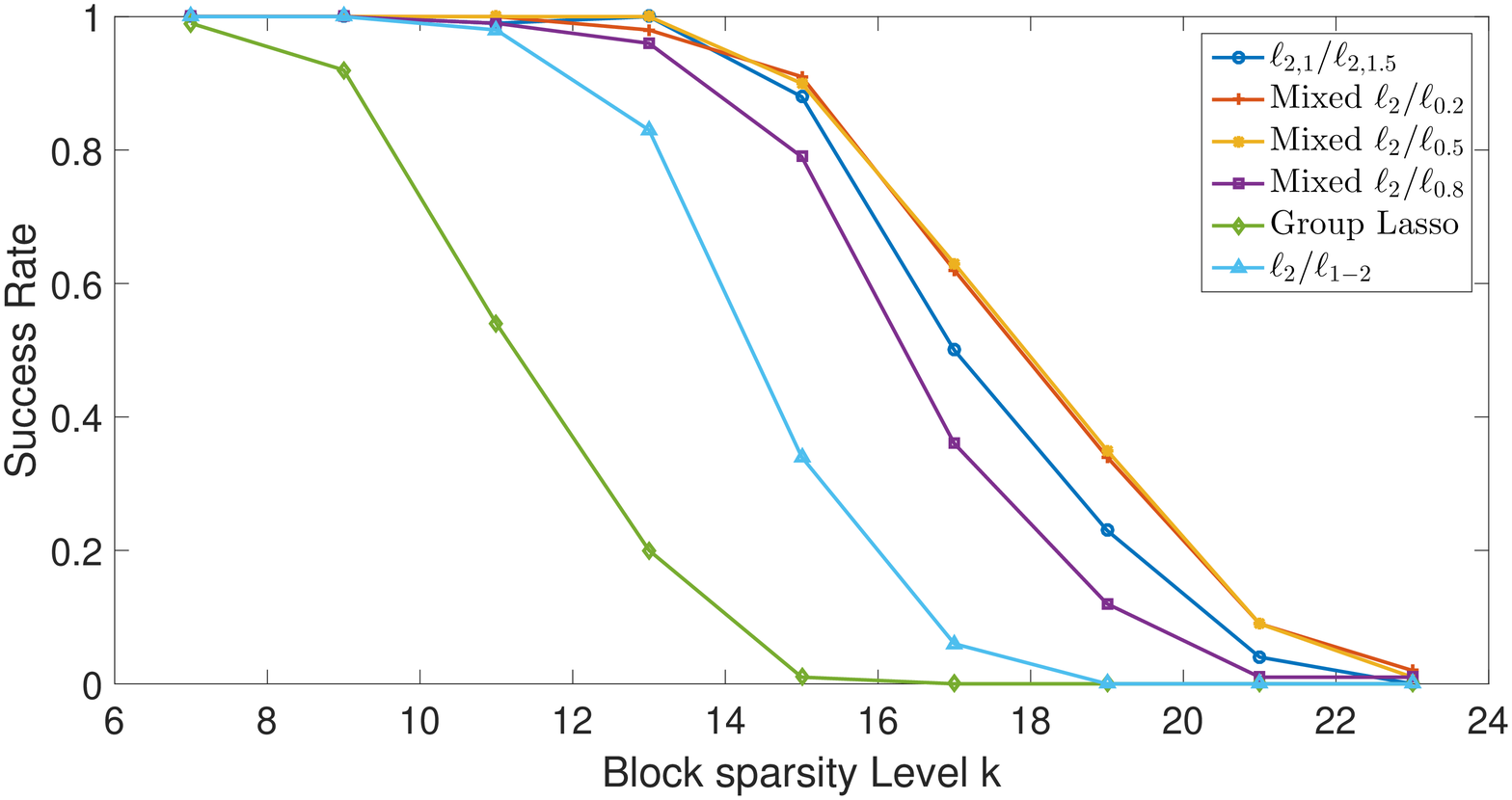}
	\caption{Recovery performance comparison for different algorithms with Gaussian random matrices.} \label{comparison_gaussian}
\end{figure}

As Figure \ref{comparison_gaussian} shows, the mixed $\ell_2/\ell_{0.2}$ and mixed $\ell_2/\ell_{0.5}$ perform the best for the Gaussian case, while our proposed $\ell_{2,1}/\ell_{2,1.5}$ tends to perform better than other methods including the mixed $\ell_2/\ell_{0.8}$, group lasso and $\ell_2/\ell_{1-2}$.

\subsubsection{Block-coherent random matrices}

Lastly, we conducted experiments with the block-coherent random matrices to verify the advantageous performance of our block $q$-ratio sparsity minimization based method in block sparse signal reconstruction. We construct the highly block-coherent random matrices $A$ as done in \cite{wang2017block} by using $A=P\otimes D$ with $D=H/\sqrt{d}$ and $P\in\mathbb{R}^{m/d\times N/d}$ is a randomly oversampled partial discrete cosine transform (DCT) matrix with its $i$-th column being $\sqrt{\frac{d}{m}}\cos(2\pi\omega (i-1)/F), i=1,2,\cdots,N/d$ and $\omega$ is a random vector uniformly distributed in $[0,1]^{m/d}$. As shown in \cite{wang2017block}, a larger $F$ yields a more block-coherent matrix. In this set of experiments, we set $m=60$, $N=500$, $d=2$, $F=5$ and $k=\{3,5,7,9,11,13,15,17\}$. 

\begin{figure}[htbp]
	\centering
	\includegraphics[width=\textwidth,height=0.4\textheight]{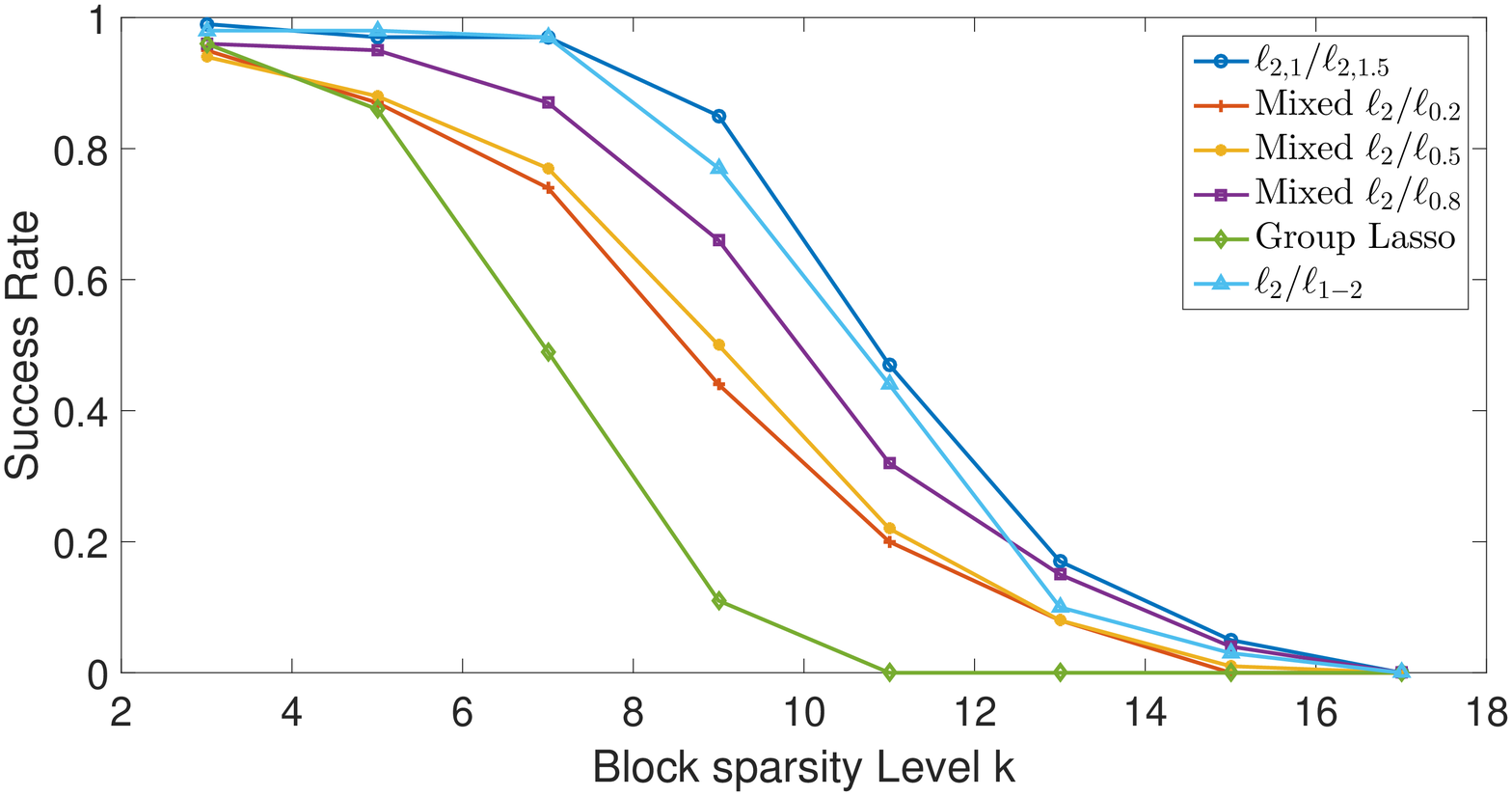}
	\caption{Recovery performance comparison for different algorithms with block-coherent random matrices.} \label{comparison_DCT}
\end{figure}

In contrast to earlier findings for the Gaussian case, however, from Figure \ref{comparison_DCT} we can see that the proposed $\ell_{2,1}/\ell_{2,1.5}$ minimization gives the best result for the block-coherent case, even better than the $\ell_2/\ell_{1-2}$ method. Taken together, these findings suggest that the $\ell_{2,1}/\ell_{2,1.5}$ can achieve satisfactory block sparse recovery results which is robust to the block coherence of the measurement matrix.

\section{Conclusion}

In this paper, we studied the block sparse signal recovery approach via minimizing the block $q$-ratio sparsity. In the case $1<q\leq \infty$, it reduces to a problem of minimizing the ratio of the mixed $\ell_2/\ell_1$ and the mixed $\ell_2/\ell_q$ norms. We gave a verifiable sufficient condition for the exact block sparse recovery and established the corresponding reconstruction error bounds in terms of $q$-ratio BCMSV. Both constrained and unconstrained models were considered. A computational algorithm was proposed to approximately solve this non-convex problem. In addition, varieties of numerical experiments were conducted to illustrate the good performance of our proposed approach. 

\bibliographystyle{apalike}
\bibliography{blocknorm_ratio}

\begin{thebibliography}{}

\bibitem[Baraniuk et~al., 2010]{baraniuk2010model}
Baraniuk, R.~G., Cevher, V., Duarte, M.~F., and Hegde, C. (2010).
\newblock Model-based compressive sensing.
\newblock {\em IEEE Transactions on Information Theory}, 56(4):1982--2001.

\bibitem[Blumensath and Davies, 2009]{blumensath2009sampling}
Blumensath, T. and Davies, M.~E. (2009).
\newblock Sampling theorems for signals from the union of finite-dimensional
  linear subspaces.
\newblock {\em IEEE Transactions on Information Theory}, 55(4):1872--1882.

\bibitem[Chartrand and Staneva, 2008]{chartrand2008restricted}
Chartrand, R. and Staneva, V. (2008).
\newblock Restricted isometry properties and nonconvex compressive sensing.
\newblock {\em Inverse Problems}, 24(3):035020.

\bibitem[Chen and Huo, 2006]{chen2006theoretical}
Chen, J. and Huo, X. (2006).
\newblock Theoretical results on sparse representations of multiple-measurement
  vectors.
\newblock {\em IEEE Transactions on Signal processing}, 54(12):4634--4643.

\bibitem[Cohen et~al., 2009]{cohen2009compressed}
Cohen, A., Dahmen, W., and DeVore, R. (2009).
\newblock Compressed sensing and best $k$-term approximation.
\newblock {\em Journal of the American Mathematical Society}, 22(1):211--231.

\bibitem[Donoho, 2006]{donoho2006compressed}
Donoho, D.~L. (2006).
\newblock Compressed sensing.
\newblock {\em IEEE Transactions on Information Theory}, 52(4):1289--1306.

\bibitem[Eldar et~al., 2010]{eldar2010block}
Eldar, Y.~C., Kuppinger, P., and Bolcskei, H. (2010).
\newblock Block-sparse signals: Uncertainty relations and efficient recovery.
\newblock {\em IEEE Transactions on Signal Processing}, 58(6):3042--3054.

\bibitem[Eldar and Kutyniok, 2012]{eldar2012compressed}
Eldar, Y.~C. and Kutyniok, G. (2012).
\newblock {\em Compressed sensing: theory and applications}.
\newblock Cambridge University Press.

\bibitem[Eldar and Mishali, 2009]{eldar2009robust}
Eldar, Y.~C. and Mishali, M. (2009).
\newblock Robust recovery of signals from a structured union of subspaces.
\newblock {\em IEEE Transactions on Information Theory}, 55(11):5302--5316.

\bibitem[Elhamifar and Vidal, 2012]{elhamifar2012block}
Elhamifar, E. and Vidal, R. (2012).
\newblock Block-sparse recovery via convex optimization.
\newblock {\em IEEE Transactions on Signal Processing}, 60(8):4094--4107.

\bibitem[Fan and Li, 2001]{fan2001variable}
Fan, J. and Li, R. (2001).
\newblock Variable selection via nonconcave penalized likelihood and its oracle
  properties.
\newblock {\em Journal of the American Statistical Association},
  96(456):1348--1360.

\bibitem[Foucart and Lai, 2009]{foucart2009sparsest}
Foucart, S. and Lai, M.-J. (2009).
\newblock Sparsest solutions of underdetermined linear systems via
  $\ell_q$-minimization for $0<q\leq 1$.
\newblock {\em Applied and {C}omputational {H}armonic {A}nalysis},
  26(3):395--407.

\bibitem[Foucart and Rauhut, 2013]{foucart2013mathematical}
Foucart, S. and Rauhut, H. (2013).
\newblock {\em A {M}athematical {I}ntroduction to {C}ompressive {S}ensing},
  volume~1.
\newblock Birkh{\"a}user Basel.

\bibitem[Grant and Boyd, 2014]{grant2014cvx}
Grant, M. and Boyd, S. (2014).
\newblock Cvx: Matlab software for disciplined convex programming, version 2.1.

\bibitem[Huang and Tran, 2018]{huang2018sparse}
Huang, S. and Tran, T.~D. (2018).
\newblock Sparse signal recovery via generalized entropy functions
  minimization.
\newblock {\em IEEE Transactions on Signal Processing}, 67(5):1322--1337.

\bibitem[Lipp and Boyd, 2016]{lipp2016variations}
Lipp, T. and Boyd, S. (2016).
\newblock Variations and extension of the convex--concave procedure.
\newblock {\em Optimization and Engineering}, 17(2):263--287.

\bibitem[Majumdar and Ward, 2010]{majumdar2010compressed}
Majumdar, A. and Ward, R.~K. (2010).
\newblock Compressed sensing of color images.
\newblock {\em Signal Processing}, 90(12):3122--3127.

\bibitem[Mishali and Eldar, 2009]{mishali2009blind}
Mishali, M. and Eldar, Y.~C. (2009).
\newblock Blind multiband signal reconstruction: Compressed sensing for analog
  signals.
\newblock {\em IEEE Transactions on Signal Processing}, 57(3):993--1009.

\bibitem[Parvaresh et~al., 2008]{parvaresh2008recovering}
Parvaresh, F., Vikalo, H., Misra, S., and Hassibi, B. (2008).
\newblock Recovering sparse signals using sparse measurement matrices in
  compressed dna microarrays.
\newblock {\em IEEE Journal of Selected Topics in Signal Processing},
  2(3):275--285.

\bibitem[Rahimi et~al., 2019]{rahimi2018scale}
Rahimi, Y., Wang, C., Dong, H., and Lou, Y. (2019).
\newblock A scale invariant approach for sparse signal recovery.
\newblock {\em SIAM Journal on Scientific Computing}, 41(6):A3649--A3672.

\bibitem[Schaible, 1976]{schaible1976minimization}
Schaible, S. (1976).
\newblock Minimization of ratios.
\newblock {\em Journal of Optimization Theory and Applications},
  19(2):347--352.

\bibitem[Schaible and Shi, 2004]{schaible2004recent}
Schaible, S. and Shi, J. (2004).
\newblock Recent developments in fractional programming: single-ratio and
  max-min case.
\newblock {\em Nonlinear analysis and convex analysis}, 493506.

\bibitem[Stancu-Minasian, 2012]{stancu2012fractional}
Stancu-Minasian, I.~M. (2012).
\newblock {\em Fractional programming: theory, methods and applications},
  volume 409.
\newblock Springer Science \& Business Media.

\bibitem[Wang et~al., 2020]{wang2020accelerated}
Wang, C., Yan, M., Rahimi, Y., and Lou, Y. (2020).
\newblock Accelerated schemes for the $ {L}_1/{L}_2 $ minimization.
\newblock {\em IEEE Transactions on Signal Processing}, 68:2660--2669.

\bibitem[Wang et~al., 2019]{wang2019error}
Wang, J., Zhou, Z., and Yu, J. (2019).
\newblock Error bounds of block sparse signal recovery based on q-ratio block
  constrained minimal singular values.
\newblock {\em EURASIP Journal on Advances in Signal Processing}, 2019(1):57.

\bibitem[Wang et~al., 2017]{wang2017block}
Wang, W., Wang, J., and Zhang, Z. (2017).
\newblock Block-sparse signal recovery via $\ell_2/\ell_{1-2}$ minimisation
  method.
\newblock {\em IET Signal Processing}, 12(4):422--430.

\bibitem[Wang et~al., 2013]{wang2013recovery}
Wang, Y., Wang, J., and Xu, Z. (2013).
\newblock On recovery of block-sparse signals via mixed $\ell_2/\ell_q(0<q\leq
  1)$ norm minimization.
\newblock {\em EURASIP Journal on Advances in Signal Processing}, 2013(1):76.

\bibitem[Wang et~al., 2014]{wang2014restricted}
Wang, Y., Wang, J., and Xu, Z. (2014).
\newblock Restricted $p$-isometry properties of nonconvex block-sparse
  compressed sensing.
\newblock {\em Signal Processing}, 104:188--196.

\bibitem[Yin et~al., 2015]{yin2015minimization}
Yin, P., Lou, Y., He, Q., and Xin, J. (2015).
\newblock Minimization of $\ell_{1-2}$ for compressed sensing.
\newblock {\em SIAM Journal on Scientific Computing}, 37(1):A536--A563.

\bibitem[Yuan and Lin, 2006]{yuan2006model}
Yuan, M. and Lin, Y. (2006).
\newblock Model selection and estimation in regression with grouped variables.
\newblock {\em Journal of the Royal Statistical Society: Series B (Statistical
  Methodology)}, 68(1):49--67.

\bibitem[Zeinalkhani and Banihashemi, 2015]{zeinalkhani2015iterative}
Zeinalkhani, Z. and Banihashemi, A.~H. (2015).
\newblock Iterative reweighted $\ell_2/\ell_1$ recovery algorithms for
  compressed sensing of block sparse signals.
\newblock {\em IEEE Transactions on Signal Processing}, 63(17):4516--4531.

\bibitem[Zhang, 2010]{zhang2010nearly}
Zhang, C.-H. (2010).
\newblock Nearly unbiased variable selection under minimax concave penalty.
\newblock {\em Annals of Statistics}, 38(2):894--942.

\bibitem[Zhang and Xin, 2018]{zhang2018minimization}
Zhang, S. and Xin, J. (2018).
\newblock Minimization of transformed ${L}_1$ penalty: theory, difference of
  convex function algorithm, and robust application in compressed sensing.
\newblock {\em Mathematical Programming}, 169(1):307--336.

\bibitem[Zhou and Yu, 2017]{zhou2017estimation}
Zhou, Z. and Yu, J. (2017).
\newblock Estimation of block sparsity in compressive sensing.
\newblock {\em arXiv preprint arXiv:1701.01055}.

\bibitem[Zhou and Yu, 2019a]{zhou2018q}
Zhou, Z. and Yu, J. (2019a).
\newblock On $ q $-ratio {CMSV} for sparse recovery.
\newblock {\em Signal Processing}, 165:128--132.

\bibitem[Zhou and Yu, 2019b]{zhou2018sparse}
Zhou, Z. and Yu, J. (2019b).
\newblock Sparse recovery based on $q$-ratio constrained minimal singular
  values.
\newblock {\em Signal Processing}, 155:247--258.

\bibitem[Zhou and Yu, 2020]{zhou2020minimization}
Zhou, Z. and Yu, J. (2020).
\newblock Minimization of the $ q $-ratio sparsity with $1< q\leq\infty $ for
  signal recovery.
\newblock {\em arXiv preprint arXiv:2010.03402}.

\end{thebibliography}

\end{document}